\documentclass[a4paper,11pt]{article}
\pdfoutput=1
\usepackage{import}
\usepackage{jheppub}
\usepackage{hyperref}
\usepackage{natbib}
\usepackage{longtable}

\usepackage{amsmath,bm,amssymb,amsthm,mathrsfs,mathtools,slashed}
\usepackage{subcaption,color}
\usepackage{ragged2e}
\usepackage{caption}
\justifying\let\raggedright\justifying

\allowdisplaybreaks[4]
\usepackage{tikz}
\usepackage{chngcntr}
\usepackage{multirow}
\usepackage{comment}
\usetikzlibrary{backgrounds}
\usetikzlibrary{shapes,matrix,trees}
\usetikzlibrary{arrows.meta}
\usepackage{centernot}
\usetikzlibrary{positioning}			
\usetikzlibrary{calc,through}			
\usetikzlibrary{decorations.pathreplacing}     
\usepackage{pgffor}                                        
\usetikzlibrary{decorations.pathmorphing}	
\usetikzlibrary{decorations.markings}
\tikzset{
	vector/.style={decorate, decoration={snake}, draw},
	provector/.style={decorate, decoration={snake,amplitude=2.5pt}, draw},
	antivector/.style={decorate, decoration={snake,amplitude=-2.5pt}, draw},
	fermion/.style={draw=black, postaction={decorate},
		decoration={markings,mark=at position .55 with {\arrow[draw=black]{>}}}},
	fermionbar/.style={draw=black, postaction={decorate},
		decoration={markings,mark=at position .55 with {\arrow[draw=black]{<}}}},
	fermionnoarrow/.style={draw=black},
	gluon/.style={decorate, draw=black,
		decoration={coil,amplitude=4pt, segment length=5pt}},
	scalar/.style={dashed,draw=black, postaction={decorate},
		decoration={markings,mark=at position .55 with {\arrow[draw=black]{>}}}},
	scalarbar/.style={dashed,draw=black, postaction={decorate},
		decoration={markings,mark=at position .55 with {\arrow[draw=black]{<}}}},
	scalarnoarrow/.style={dashed,draw=black},
	electron/.style={draw=black, postaction={decorate},
		decoration={markings,mark=at position .55 with {\arrow[draw=black]{>}}}},
	bigvector/.style={decorate, decoration={snake,amplitude=4pt}, draw},
	photon/.style={decorate, draw=black,decoration={snake,amplitude=4pt, segment length=5pt} }
}

\definecolor{ccblue}{rgb}{0.0,0.4,0.8}
\usepackage[]{hyperref}
\hypersetup{  colorlinks=true,
	linkcolor=ccblue,
	urlcolor=ccblue,
	citecolor=ccblue}
\usepackage{amsmath}
\usepackage{amsfonts}
\usepackage{amssymb}
\usepackage{booktabs}
\usepackage{slashed}
\usepackage[]{hyperref}
\usepackage{autobreak}
\graphicspath{ {figs/}}

\newcommand{\nn}{\nonumber}
\newcommand{\ep}{\epsilon}
\newcommand{\lih}{{\rm Li}_4 ({\scriptstyle \frac{1}{2}})}

\newcommand{\lifh}{{\rm Li}_5 ({\scriptstyle \frac{1}{2}})}

\newcommand{\iomx}{\bar{x}}
\newcommand{\fxmo}{f_{\scriptscriptstyle -1}~}
\newcommand{\fxmosq}{f_{\scriptscriptstyle -1}^{2}~}

\newcommand{\fxei}{f_{\scriptscriptstyle 8}~}
\newcommand{\fxtnt}{f_{\scriptscriptstyle 2,3}~}

\newcommand{\ixthr}{x^{{\scriptscriptstyle -}3}}
\newcommand{\ixtw}{x^{{\scriptscriptstyle -}2}}
\newcommand{\ixo}{x^{{\scriptscriptstyle -}1}}

\newcommand{\asr}{\Big( \frac{\alpha_s}{4\pi} \Big)}
\newcommand{\asrL}{\Big( \frac{\bar{\alpha}_s}{4\pi} \Big)}
\newcommand{\D}{{\cal D}}

\title{Three loop QCD corrections to the heavy-light form factors: 
fermionic contributions}

\author[a]{Sudeepan Datta,}
\author[b]{Narayan Rana}

\affiliation[a]{Centre for High Energy Physics, Indian Institute of Science, Bangalore 560012, India}
\affiliation[b]{School of Physical Sciences, National Institute of Science Education and Research,\\ An OCC of Homi Bhabha National Institute, Jatni 752050, India}

\emailAdd{sudeepand@iisc.ac.in, narayan.rana@niser.ac.in}

\abstract{
We present analytic results for three-loop fermionic corrections to
the heavy-light form factors in perturbative quantum chromodynamics.
Specifically, we present all light quark contributions and contributions 
from two heavy quark loops.
We use the method of differential equations to compute all relevant 
three-loop master integrals. The results for all these contributions are expressed in terms of harmonic polylogarithms and generalized harmonic polylogarithms.
}

\newpage

\begin{document} 
\maketitle
\counterwithin{table}{section}

\section{Introduction}
The basic ingredients of scattering amplitudes in quantum field 
theories are the amplitudes involving two on-shell states of elementary fields and an off-shell state described through a composite operator,
otherwise known as the form factors.
Essential information about the analytic structure of generic scattering 
amplitudes can be obtained from rigorous studies of the form factors. They are also important objects of study in the context of theoretical precision calculations for collider phenomenology, since 
major contributions to the physical observables at the colliders such as scattering cross-sections and decay rates 
come from virtual amplitudes, where the latter 
themselves are written as a combination of the form factors.
The form factors with two massless external particles (gluons or massless quarks) coupled to an external current, play a critical role in 
the Standard Model (SM) precision studies. Hence, they have been obtained
up to four loops in perturbative quantum chromodynamics (pQCD) \cite{Ravindran:2004mb,deFlorian:2013sza,Moch:2005tm,Moch:2005id,Baikov:2009bg,Gehrmann:2010ue,Gehrmann:2014vha,Ahmed:2015qia,Ahmed:2015qpa,Ahmed:2016vgl,Ahmed:2016qjf,Ahmed:2019yjt,Lee:2021lkc,Lee:2022nhh,Chakraborty:2022yan}.
The heavy quark form factors (HQFF) are the cases when
both the external states are massive quarks with the same mass.
Because of their high importance in the study of the forward-backward asymmetry of heavy quark pair production at lepton colliders, 
the anomalous magnetic moment of a heavy quark, and the top-quark properties at the High-Luminosity Large Hadron Collider (HL-LHC),
the HQFF have been computed up to three loops in pQCD  
\cite{Bernreuther:2004ih,Bernreuther:2004th,Bernreuther:2005rw,Bernreuther:2005gw,Gluza:2009yy,Ablinger:2017hst,Henn:2016tyf,Lee:2018nxa,Ablinger:2018yae,Lee:2018rgs,Ablinger:2018zwz,Blumlein:2019oas,Fael:2022rgm,Fael:2022miw,Fael:2023zqr,Blumlein:2023uuq}.

The heavy-light form factors (HLFF) are defined for a pair of heavy and light external quarks. They are essential in studying flavour-changing decays of heavy quarks, such as a top quark ($t$) 
decaying into a bottom quark ($b$) and a charged $W$ boson.
The analytically continued HLFF constitute most of the virtual corrections for such decays. 
At present, a precise determination of the top quark decay width ($\Gamma_t$) is highly 
demanding and hence has been obtained up to next-to-next-to-next-to leading order (N$^3$LO) in pQCD \cite{Chetyrkin:1999ju,Blokland:2004ye,Gao:2012ja,Brucherseifer:2013iv,Chen:2023dsi,Chen:2023osm,Yan:2024hbz}. 
The infrared (IR) subtracted HLFF for $b \rightarrow u W^{*}$ 
are also essential in computing hard functions to study the inclusive semileptonic decay $B \rightarrow X_u\, l\, \bar{\nu_l}$ in the shape-function region, using Soft Collinear Effective Theory (SCET). Such decays play a vital role in studying flavour and CP-violation in the quark sector by precisely determining the Cabibbo–Kobayashi–Maskawa (CKM) matrix element $|V_{ub}|$. Recent reviews on inclusive semileptonic $B$-decays and allied aspects can be found in \cite{Gambino:2020jvv,Bharucha:2022zci,Fael:2024rys,Mandal:2024lpl}.

The explicit computation of the two-loop HLFF considering the light quark to be massless, has been discussed 
in refs.~\cite{Bonciani:2008wf,Huber:2009se,Bell:2006tz}, in the pQCD framework.
The full dependence of the two-loop HLFF on the light quark mass has also been computed in 
refs.~\cite{Chen:2018dpt,Engel:2018fsb}.
A first step towards N$^3$LO predictions has been taken by computing the master integrals (MIs) appearing in the color-planar contributions at three loops considering the light quark to be massless in ref.~\cite{Chen:2018fwb}. 
In ref.~\cite{Datta:2023otd}, we have obtained the first results for the form factors considering the light 
quark as massless, in the color-planar limit.
Very recently, during the finalization of the present work, the complete form factors, also with the 
approximation of zero light quark mass, has been 
presented using a semi-analytic approach in ref.~\cite{Fael:2024vko}.

In this work, we present analytic results for all light quark contributions and contributions from diagrams with two heavy quark loops to the three-loop QCD corrections to the HLFF.
In line with our previous work for the color-planar contributions, we maintain the light quark to be massless.
We follow the standard procedure of reducing all Feynman integrals 
to a basis of MIs using the integration-by-parts (IBP) reduction technique \cite{Chetyrkin:1981qh,Laporta:2001dd},
and solving the MIs analytically using the method of differential equations \cite{Kotikov:1990kg,Argeri:2007up,Remiddi:1997ny,Henn:2013pwa,Ablinger:2015tua}. 
The ultraviolet (UV) renormalization of the fields and masses are performed in a mixed scheme. We also show that the universal IR behaviour \cite{Becher:2009kw} is satisfied by the UV renormalized form factors.

This paper is organized as follows. 
We briefly discuss the theoretical setup in section \ref{sec:amps}, and the details of the 
computation in section \ref{sec:comp}.
In section \ref{sec:results}, we present the primary result
of this paper i.e. the three-loop finite remainders for all light quark contributions and contributions 
from diagrams with two heavy quark loops, 
after performing appropriate IR subtraction. 
In section \ref{sec:conclusions}, we present our concluding remarks.

\section{Theoretical framework}
\label{sec:amps}
The HLFF are matrix elements of an external current between a pair of heavy 
and light quark states. 
We consider the decay of a top quark ($t$) into a massless 
bottom quark ($b$) and a off-shell $W$ boson, as a physical process
which is given by
\begin{equation}
    t (P) \rightarrow  b (p) + W^* (q) \,.
\end{equation}
We define the momenta of the incoming top and the outgoing bottom quark
to be $P$ and $p$, respectively.
From the momentum conservation, the momentum of 
the $W$ boson is $q = P-p$. 
Both the quarks are on-shell, which can be written as
\begin{equation}
    P^2 = m_t^2\,, ~~ p^2 = 0 \,,
\end{equation}
where, $m_t$ is the mass of the top quark. 
We introduce $x$, a dimensionless variable, as
    $x = \frac{q^2}{m_t^2} \,$.
A detailed description of the generic structure of the amplitude,
along with the definitions of the form factors and corresponding 
projectors have been presented in \cite{Datta:2023otd}.
For the purpose of defining the form factors $G_1, G_2, G_3$, we introduce the 
generic form of the vertex 
with one massive and one massless quark as
\begin{equation}
 \Gamma^{\mu}_{cd} = -i \delta_{cd} \Big[ G_1 \gamma^\mu (1-\gamma_5) + \frac{G_2}{2 m_t} (1+\gamma_5) ( P^\mu + p^\mu )
    + \frac{G_3}{2 m_t} (1+\gamma_5) ( P^\mu - p^\mu ) \Big] \,.
\end{equation}
$\Gamma^{\mu}_{cd}$ with the bi-spinors $t_d (P)$ and 
$\bar{b}_c (p)$ of $t$-quark and $b$-quark, respectively, 
constitute the general form of the amplitude as
\begin{equation}
 \bar{b}_c (p) ~ \Gamma^{\mu}_{cd} ~ t_d (P) \,.
\end{equation}
Similarly, we define the pseudo-scalar vertex which denotes the 
process where a top quark decays into a bottom quark and a charged 
pseudo-Goldstone boson, as follows
\begin{equation}
 \Gamma_{\text{PS}} = \frac{m_t}{m_W} S (1+\gamma_5) \,,
\end{equation}
where $S$ is the pseudo-scalar form factor.
The form factors are computed using pQCD and
admit a series expansion in powers of the strong coupling constant ($\alpha_s$) as 
\begin{equation}
    G_i = \frac{i g_w}{2 \sqrt{2}} \sum_{n=0}^\infty \Big( \frac{\alpha_s}{4\pi} \Big)^n G_i^{(n)} \,,
\qquad
\qquad
 S = \frac{i g_w}{2 \sqrt{2}} \sum_{n=0}^\infty \Big( \frac{\alpha_s}{4\pi} \Big)^n S^{(n)} \,.
\end{equation}
$g_w = \frac{e}{s_w}$
where $e$ is the electric charge and $s_w$ denotes the sine of the weak mixing angle.
The leading order (LO) form factors are
\begin{equation}
    G_1^{(0)} = 1, \quad G_2^{(0)} = 0, \quad G_3^{(0)} = 0,
    \quad S^{(0)} = 1.
\end{equation}

The bare form factors contain divergences of UV and IR origin.
We use dimensional regularization in $d=4-2\epsilon$ space-time dimensions to regulate these divergences. 
Because of the presence of $\gamma_5$, which is inherently a 
four-dimensional object, it is mandatory to choose a scheme
to treat $\gamma_5$ in $d$ space-time dimensions. 
As both the appearing $\gamma_5$, one from the vertex and the 
other from the projectors, are connected to open fermion lines, 
we use the prescription presented in \cite{Kreimer:1989ke,Korner:1991sx} i.e., the anti-commutation of $\gamma_5$ with other $\gamma$-matrices (\{$\gamma_\mu,\gamma_5\} = 0$), followed by $\gamma_5^2 = 1$.

To cancel the divergences of UV origin, we perform the UV 
renormalization of the fields and parameters, following a 
mixed scheme. The wave function ($Z_{2,t}$) and the mass ($Z_m$)
of the heavy quark ($t$) have been renormalized using the on-shell (OS) 
renormalization scheme 
as well the wave function ($Z_{2,b}$) of the massless quark ($b$) and the mass of the heavy quark appearing in the pseudo-scalar vertex.
For the strong coupling constant ($Z_{\alpha_s}$), we use the modified minimal subtraction ($\overline{\text{MS}}$) scheme \cite{tHooft:1973mfk,Bardeen:1978yd}.
All the renormalization constants admit a series expansion in $\alpha_s$ as 
\begin{equation}
 Z_i = \sum_{n=0}^\infty \Big( \frac{\alpha_s}{4 \pi} \Big)^n Z_i^{(n)} \,. 
\end{equation}
We require $Z_m$ \cite{Broadhurst:1991fy,
Melnikov:2000zc,Marquard:2007uj,Marquard:2015qpa,Marquard:2016dcn},
and $Z_{2,t}$ \cite{Broadhurst:1991fy, Melnikov:2000zc,Marquard:2007uj,Marquard:2018rwx} 
up to three loops in QCD.
$Z_{\alpha_s}$ is needed only up to two loops, which are available in refs.~\cite{Tarasov:1980au,Larin:1993tp,vanRitbergen:1997va,Czakon:2004bu,Baikov:2016tgj,Herzog:2017ohr,Luthe:2017ttg}.
The two-loop results for $Z_{2,b}$ have been presented in
ref.~\cite{Bonciani:2008wf,Beneke:2008ei,Bell:2008ws}, while the three-loop results can be found in \cite{Grozin:2011nk,Gerlach:2018hen}.

The renormalization of the quark wave functions, the strong coupling constant
and the Yukawa coupling is multiplicative. While the renormalization of the mass ($m_t$)
of the heavy quark is performed by explicitly computing the Feynman diagrams 
with counter-terms inserted. 
Schematically, the renormalized form factors ($G_i$)
can be written in terms of the bare form factors ($\hat{G}_i$) as follows
\begin{equation}
 G_i = Z_{2,t}^{\frac{1}{2}} Z_{2,b}^{\frac{1}{2}} ( \hat{G}_i + \hat{G}_{ct,i} )\,.
\end{equation}
The counter-term contributions from heavy quark mass renormalization is denoted 
by $\hat{G}_{ct,i}$. 
We consider the strong coupling constant in full QCD with $n_l$ and $n_h$ number of massless and massive quarks, respectively.
Correspondingly, the $\beta$-functions up to two loops are given by
\begin{align}
 \beta_0 &= \frac{11}{3} C_A - \frac{4}{3} T_F (n_h + n_l) \,,
 \nonumber\\
 \beta_1 &= \frac{34}{3} C_A^2 - \frac{20}{3} C_A T_F (n_h+n_l) - 4 C_F T_F (n_h+n_l) \,.  
\end{align}
$C_A = N_C$ and $C_F = \frac{N_C^2-1}{2N_C}$ are the adjoint and fundamental 
Casimirs, respectively, of the SU($N_C$) gauge theory. 
The on-shell Ward identity connects the two vertices as follows
\begin{equation}
 q_\mu \Gamma^\mu - m_W \Gamma_{\text{PS}} = 0 \,.
\end{equation}
The identity can be written in terms of the form factors as follows
\begin{equation}
\label{eq:WI}
 2 G_1^{(n)} + G_2^{(n)} + x G_3^{(n)} - 2 S^{(n)} = 0 \,.
\end{equation}

%

IR divergences arising from soft gluons and collinear partons, of the
UV renormalized scattering amplitudes in QCD manifest a universal behaviour.
In case of massless scattering amplitudes, the universal structure
has been studied in great detail in \cite{Catani:1998bh,Sterman:2002qn,Becher:2009cu,Gardi:2009qi,Ravindran:2006cg}. For the scattering amplitudes with massive external particles, 
the general IR structure was developed in \cite{Mitov:2006xs,Ahmed:2017gyt,Blumlein:2018tmz}
in the high-energy limit.
In ref.~\cite{Becher:2009kw}, 
a generic scattering amplitude with both massless and massive external particles has been considered,
and corresponding universal IR structure has been presented at two-loop level. 
We follow ref.~\cite{Becher:2009kw} to write the IR structure of the HLFF as a 
multiplicative factor as follows
\begin{equation}
 G_i = Z(\bar{\mu}) G_i^{\text{fin}} (\bar{\mu}) \,.
 \label{eq:ir}
\end{equation}
$G_i^{\text{fin}} (\bar{\mu})$ is finite as $\ep \rightarrow 0$.
We introduce the scale $\bar{\mu}$ for this particular factorization of IR singularities.
The discussion on the dependence of $Z(\bar{\mu})$ on the corresponding anomalous dimension ($\Gamma$)
through the decoupling relation of $\alpha_s$ has been discussed in detail in \cite{Blumlein:2019oas,Datta:2023otd}.
While $Z$ is the universal IR multiplicative factor in full QCD, an equivalent factor
$\bar{Z}$ can be introduced for massless QCD, the solution for which can be found up to three loops
in terms of the anomalous dimension and $\beta$-function following the renormalization group equation as
\begin{align}
 \ln \bar{Z} &= \asrL \bigg[ \frac{\Gamma_0'}{4 \ep^2}  + \frac{\Gamma_0}{2 \ep} \bigg]
       + \asrL^2 \bigg[ - \frac{3 \bar{\beta}_0 \Gamma_0'}{16 \ep^3} + \frac{\Gamma_1' - 4 \bar{\beta}_0 \Gamma_0}{16 \ep^2} + \frac{\Gamma_1}{4 \ep} \bigg]
       \nn\\
       &+ \asrL^3 \bigg[ \frac{11 \bar{\beta}_0^2 \Gamma_0'}{72 \ep^4} - \frac{5 \bar{\beta}_0 \Gamma_1' 
       + 8 \bar{\beta}_1 \Gamma_0' - 12 \bar{\beta}_0^2 \Gamma_0}{72 \ep^3}
                      + \frac{\Gamma_2' - 6 \bar{\beta}_0 \Gamma_1 - 6 \bar{\beta}_1 \Gamma_0}{36 \ep^2} + \frac{\Gamma_2}{6 \ep} \bigg]
       \nn\\
       &+ {\cal O}(\bar{\alpha}_s^4) \,,
\end{align}
where, $\Gamma$ is expanded in perturbative series in $\bar{\alpha}_s$
\begin{equation}
 \Gamma = \sum_{n=0}^\infty \asrL^{n+1} \Gamma_n \,,
 \quad 
 \textmd{and}
 \quad 
  \Gamma_n' = \frac{\partial}{\partial \bar{\mu}} \Gamma_n \,.
\end{equation}
In case of HLFF, $\Gamma$ receives contributions from massless and massive cusp anomalous dimensions \cite{Korchemsky:1987wg,Korchemsky:1991zp,Grozin:2014hna,Grozin:2015kna}.
$\bar{\beta}$ is the $\beta$-function in the massless QCD.
The decoupling relation of the strong coupling 
constant \cite{Chetyrkin:1997un,Grozin:2007fh,Grozin:2011nk,Schroder:2005hy}
can be used to obtain $Z$ from $\bar{Z}$.

\section{Computational details}
\label{sec:comp}
\addtocounter{table}{-2}
The computational technique adopted to obtain the three-loop HLFF in this paper
is the conventional method of generating Feynman diagrams using \textsc{QGRAF} \cite{Nogueira:1991ex},
followed by in-house \textsc{FORM} \cite{Tentyukov:2007mu} routines to 
convert the \textsc{QGRAF} output into Feynman amplitudes and for further manipulations
of Dirac, Lorentz and color algebra.
We have also used \textsc{Color} \cite{vanRitbergen:1998pn} to perform the color algebra.
The total number of diagrams present at two- and three-loop levels are 13 and 263, respectively.

The expressions after the manipulations of all the algebras contain a large number of scalar Feynman integrals. Following the standard method of IBP reduction, we reduce these scalar Feynman integrals
to a set of chosen Feynman integrals, the MIs, which are much smaller in number. 
We use the public code \textsc{Kira} \cite{Klappert:2020nbg} for IBP reduction.
An intermediate step involves choosing suitable integral families. In the next, we present the details
that are relevant for the present computation.

In this paper,
we present the complete light fermionic contributions and contributions from double heavy quark loops
to the three-loop HLFF. In terms of Casimirs, these contributions are coefficients of
$C_F n_l^2$, 
$C_F^2 n_l$, 
$C_F C_A n_l$, 
$C_F n_l n_h$,
and 
$C_F n_h^2$.
The scalar Feynman integrals for these contributions are mapped to five integral families, which are
presented in the following.
We use \textsc{Reduze} \cite{vonManteuffel:2012np} to perform this mapping.
The five integral families are
\begin{equation}
\label{Cfams}
\begin{split}
    C_1 & : \{\bar{\D}_1,\, \bar{\D}_2,\, \bar{\D}_3,\, \D_{12},\, \D_{23},\, \D_{13},\,  \D_{1;1},\, \D_{2;1},\, \D_{3;1},\, \D_{1;12},\, \D_{2;12},\, \D_{3;12}\} \\
    C_2 & : \{\bar{\D}_1,\, \D_2,\, \D_3,\,  \D_{12},\, \D_{23},\, \bar{\D}_{13},\, \D_{1;1},\, \D_{2;1},\, \bar{\D}_{3;1},\,  \D_{1;12},\, \D_{2;12},\, \D_{3;12}\} \\
    C_3 & : \{\bar{\D}_1,\, \bar{\D}_2,\, \D_{12},\,  \D_{23},\, \D_{13},\, \D_{1;1},\,  \D_{3;1},\, \D_{12;2},\, \D_{2;12},\,  \D_{3;12},\, \D_3,\, \D_{2;1}\} \\
    C_4 & : \{\bar{\D}_1,\, \bar{\D}_2,\, \D_3,\,  \D_{12},\, \bar{\D}_{23},\, \bar{\D}_{13},\,  \D_{1;1},\, \D_{2;1},\, \bar{\D}_{3;1},\,  \D_{1;12},\, \D_{2;12},\, \D_{3;12}\} \\
    C_7 & : \{\bar{\D}_1,\, \D_2,\, \D_3,\, \bar{\D}_{12},\, \D_{23},\, \bar{\D}_{13},\,  \D_{1;1},\, \bar{\D}_{2;1},\, \bar{\D}_{3;1},\,  \D_{1;12},\, \D_{2;12},\, \D_{3;12}\}
\end{split}
\end{equation}
where,
\begin{equation}
 \D_i = k_i^2,\, \D_{ij} = (k_i-k_j)^2,\, \D_{i;1} = (k_i-P)^2,\, \D_{i;12} = (k_i-P+p)^2,\, \D_{ij;2}=(k_i-k_j-p)^2\,,
\end{equation}
and
\begin{equation}
 \bar{\D}_i = \D_i-m_t^2,\,
 \bar{\D}_{ij} = \D_{ij}-m_t^2,\,
 \bar{\D}_{i;1} = \D_{i;1}-m_t^2,\,
 \bar{\D}_{i;12} = \D_{i;12}-m_t^2,\,
 \bar{\D}_{ij;2} = \D_{ij;2}-m_t^2\,.
\end{equation}
To fix our notation, we present an example of generic three-loop integral as follows:
\begin{equation}
    I_{\nu}(d,x)= \int \prod_{i=1}^3 \frac{d^d k_i}{ (2 \pi)^{d}} \prod_{j=1}^{12} \frac{1}{D_{j}^{\nu_j}}
\end{equation}
where, $\nu = \nu_1\nu_2...\nu_{12}$.
We put $S_\ep = \exp ( - \ep (\gamma_E - \ln (4\pi)))$ equal to one for each loop order.

Although, five integral families are required to map all the intended Feynman diagrams, only a few sectors
of these integral families are enough to be considered for reduction. In Table~\ref{table:C2347mi}, 
we present the appearing sectors and corresponding MIs for all the integral families except $C_1$, which has been presented in \cite{Datta:2023otd}.
\begin{table}
    \begin{minipage}{0.48\textwidth}
    \begingroup
     \centering
     \begin{longtable}{ |p{0.2cm}|p{1.0cm}| p{0.65\textwidth}| }
      \hline 
       \# & sector & \hfil $C_2$ master integrals  \\  
       \hline 
       \hline 
     5  & 307 &   $I_{110011001000}$  \\
        & 818 &   $I_{010011001100}$, $I_{(-1)10011001100}$   \\
        & 1321 &   $I_{100101001010}$, $I_{1(-1)0101001010}$,   \\
        &     &    $I_{10(-1)101001010}$ \\
        & 1324 &  $I_{001101001010}$, $I_{(-1)01101001010}$   \\
     \hline  
     \hline
     6  & 819 &   $I_{110011001100}$, $I_{11(-1)011001100}$   \\
        & 937 &   $I_{100101011100}$, $I_{1(-1)0101011100}$   \\
        & 940 &   $I_{001101011100}$, $I_{(-1)01101011100}$   \\
        & 1449 &   $I_{100101011010}$   \\
        & 1452 &   $I_{001101011010}$  \\
    \hline
     \end{longtable}
    \endgroup
~
    \begingroup
     \centering
     \begin{longtable*}{ |p{0.2cm}|p{1.0cm}| p{0.65\textwidth}| }
    \hline
     \# & sector & \hfil $C_4$ master integrals  \\  
       \hline 
       \hline 
     4  & 51  &   $I_{110011000000}$  \\   
        & 275 & $I_{110010001000}$ \\
     \hline
     \hline
     5  & 803 &   $I_{110001001100}$, $I_{11(-1)001001100}$  \\
        & 307 &   $I_{011100100000}$   \\
     \hline
    \end{longtable*}
    \endgroup
    \end{minipage}
    \begin{minipage}{0.49\textwidth}
    \begingroup
     \centering
     \begin{longtable*}{ |p{0.2cm}|p{1.0cm}| p{0.65\textwidth}| }
      \hline 
       \# & sector & \hfil $C_3$ master integrals  \\  
       \hline 
       \hline
    5   & 651 &  $I_{110100010100}$, $I_{11(-1)100010100}$\\
        &     &  $I_{1101(-1)0010100}$\\
    \hline
    \hline
    6   & 411 &  $I_{110110011000}$, $I_{120110011000}$,\\
        &     &  $I_{210110011000}$, $I_{11(-1)110011000}$,\\
        &     &  $I_{11011(-1)011000}$ \\
        & 467 &  $I_{110010111000}$, $I_{11(-1)010111000}$,\\
        &     &  $I_{110(-1)10111000}$\\
        & 683 & $I_{110101010100}$, $I_{11(-1)101010100}$\\
    \hline
    7   & 443 &  $I_{110111011000}$ \\
        & 471 &  $I_{111010111000}$ \\
        & 687 &  $I_{111101010100}$ \\
    \hline
    \end{longtable*}
    \endgroup
~

~

~
    \begingroup
     \centering
     \begin{longtable*}{ |p{0.2cm}|p{1.0cm}| p{0.65\textwidth}| }
    \hline
     \# & sector & \hfil $C_7$ master integrals  \\  
       \hline 
       \hline
       6  & 937  &   $I_{100101011100}$, $I_{1(-1)0101011100}$  \\ 
     \hline
    \end{longtable*}
    \endgroup
    \end{minipage}
    \caption{List of the master integrals for $C_2$, $C_3$, $C_4$, and $C_7$. \# indicates the number 
        of propagators.}
    \label{table:C2347mi}
    \end{table}

\subsection{Computation of the master integrals}
We compute the MIs using the method of differential equations \cite{Kotikov:1990kg,Argeri:2007up,Remiddi:1997ny,Henn:2013pwa,Ablinger:2015tua}.
As usual, we differentiate the MIs with respect to $x$ and reduce the differentiated results using IBP
identities to obtain the system of differential equations.
We organize the system in an upper block-triangular form and solve it using bottom-up approach.

In this approach, the last block (corresponding to a given coupled subsystem) is homogeneously coupled, so we calculate in a bottom-up manner. We expand each block in a series in $\ep$ and solve successively for each order in $\epsilon$ starting with the leading singular term.
Furthermore, at each order of the $\epsilon$-expansion, the block is decoupled, thereby leading to higher-order differential 
equations, the operators for which are found to factorise to first-order. Thus, the relevant function space for the calculation is spanned only by multiple polylogarithms.
The full system is finally solved by using the method of variation of constants.

In this case also, just as in the color-planar scenario, the spanning alphabet is
\begin{equation}
\label{eq:LETT}
\bigg\{
\frac{1}{x},~~
\frac{1}{1-x},~~
\frac{1}{1+x},~~
\frac{1}{2-x}  \bigg\}
\end{equation}
i.e.~the usual harmonic polylogarithms (HPLs) \cite{Remiddi:1999ew} and 
generalized HPLs \cite{Ablinger:2013cf}. 
The kernel with letter $2$ is defined such that
\begin{equation}
 H_2 (x) = \ln (2) - \ln (2-x) \,.
\end{equation}
The solutions of the differential equations require boundary conditions which are 
determined at various values of $x$, namely, at $x=0$, $x=1$ or $x=-1$, or demanding regularity at $x=0$. 
We have used \textsc{MBConicHulls} \cite{Ananthanarayan:2020fhl,Banik:2022bmk,Banik:2023rrz} with \textsc{AMBRE} \cite{Gluza:2007rt}, and \textsc{HypExp} \cite{Huber:2005yg,Huber:2009se}
to compute boundary
conditions for several two-point three-loop MIs. For some cases, we have also used \textsc{MultiHypExp} \cite{Bera:2022ecn,Bera:2023pyz} for a deep expansion in $\epsilon$ of hypergeometric functions appearing after residue computation and converting the resulting series to a closed-form.
Most of the boundary conditions can be evaluated demanding regularity conditions at various values of $x$.
For a few MIs, explicit computations of the boundary conditions are required. 
We employ the method of auxiliary mass flow \cite{Liu:2017jxz,Liu:2021wks} through \textsc{AMFlow} to
obtain highly precise (400 digits) numerical output for those MIs at $x=-1$.
On the other hand, study of the differential equations and corresponding polylogarithms
indicates that the appearing constants are multiple zeta values (MZVs) \cite{Blumlein:2009cf}, 
$\ln (2)$, and ${{\rm Li}_n ({\scriptstyle \frac{1}{2}})}$.
An appropriate set of constants can be guessed for each order in $\ep$ for all the MIs,
which allows us to use the \textsc{PSLQ} algorithm \cite{pslq:92} to 
obtain the analytic expressions for the boundary conditions.

In the following, we present five MIs from $C_2$ and $C_4$ integral families,
which were needed to be computed explicitly.
Except $C_{2;2}$, the rest of the four MIs are constant integrals.
$C_{2;2}$ has been presented for $x=-1$.
The two MIs from $C_2$ integral family are
\begin{align}
 & C_{2;1} := I_{110011001000} (-1)\,, ~~ C_{2;2} := I_{100101001010} (-1)\,.
\end{align}
They are given by 
\begin{align}
 C_{2;1} =&
 \frac{1}{\ep^3} \bigg( \frac{2}{3} \bigg)
+ \frac{1}{\ep^2} \bigg( \frac{10}{3} \bigg)
+ \frac{1}{\ep} \bigg(  \frac{26}{3}
+3 \zeta_2
 \bigg)
+ \bigg( 
2
+27 \zeta_2
+\frac{14}{3} \zeta_3
\bigg)
+ \ep \bigg(  
-\frac{398}{3}
+159 \zeta_2
+\frac{287}{20} \zeta_2^2
\nn\\&
+\frac{238}{3} \zeta_3
-96 \zeta_2 \log (2)
\bigg)
+ \ep^2 \bigg( 
-1038
+512 \lih
+777 \zeta_2
+\frac{323}{20} \zeta_2^2
+\frac{1862}{3} \zeta_3
+21 \zeta_2 \zeta_3
\nn\\&
+\frac{478}{5} \zeta_5
-960 \zeta_2 \log (2)
+256 \zeta_2 \log ^2(2)
+\frac{64 \log ^4(2)}{3}
\bigg)
+ \ep^3 \bigg( 
-\frac{17470}{3}
+5120 \lih
\nn\\&
+4096 \lifh
+3435 \zeta_2
-\frac{7389}{20} \zeta_2^2
+\frac{17357}{168} \zeta_2^3
+3598 \zeta_3
+81 \zeta_2 \zeta_3
+\frac{49}{3} \zeta_3^2
-2498 \zeta_5
\nn\\&
-6144 \zeta_2 \log (2)
+\frac{2224}{5} \zeta_2^2 \log (2)
+2560 \zeta_2 \log ^2(2)
-\frac{2048}{3} \zeta_2 \log ^3(2)
+\frac{640 \log ^4(2)}{3}
\nn\\&
-\frac{512}{15} \log ^5(2)
\bigg)
+ {\mathcal{O}} (\ep^4) \,.
%
\\
 C_{2;2} =&
 \frac{1}{\ep^3} \bigg( \frac{2}{3} \bigg)
+ \frac{1}{\ep^2} \bigg( \frac{11}{3} \bigg)
+ \frac{1}{\ep} \bigg( 12
+\frac{3}{2} \zeta_2
 \bigg)
+ \bigg( 
\frac{70}{3}
+\frac{39}{2} \zeta_2
-\frac{7}{12} \zeta_3
\bigg)
+ \ep \bigg(  
-\frac{62}{3}
+\frac{291}{2} \zeta_2
\nn\\&
+\frac{1}{10} \zeta_2^2
+\frac{217}{6} \zeta_3
-90 \zeta_2 \log (2)
\bigg)
+ \ep^2 \bigg(  
-510
+360 \lih
+\frac{1701}{2} \zeta_2
-\frac{2273}{40} \zeta_2^2
+399 \zeta_3
\nn\\&
-\frac{21}{8} \zeta_2 \zeta_3
+\frac{1603}{80} \zeta_5
-945 \zeta_2 \log (2)
+270 \zeta_2 \log ^2(2)
+15 \log ^4(2)
\bigg)
+ \ep^3 \bigg(  
-\frac{10462}{3}
\nn\\&
+3780 \lih
+2880 \lifh
+\frac{8703}{2} \zeta_2
-\frac{23763}{40} \zeta_2^2
-\frac{6677}{336} \zeta_2^3
+\frac{8113}{3} \zeta_3
-\frac{969}{4} \zeta_2 \zeta_3
\nn\\&
+\frac{343}{48} \zeta_3^2
-\frac{84523}{40} \zeta_5
-6345 \zeta_2 \log (2)
+459 \zeta_2^2 \log (2)
+2835 \zeta_2 \log ^2(2)
-720 \zeta_2 \log ^3(2)
\nn\\&
+\frac{315 \log ^4(2)}{2}
-24 \log ^5(2)
\bigg)
+ {\mathcal{O}} (\ep^4) \,.
\end{align}
%
The three constant MIs from $C_4$ integral family are
\begin{align}
 & C_{4;1} := I_{110011001000}\,, ~~ C_{4;2} := I_{110011000000}\,, 
 ~~ C_{4;3} := I_{110010001000} \,,
\end{align}
which are given by
\begin{align}
 C_{4;1} =&
 \frac{1}{\ep^3} 
+ \frac{1}{\ep^2} \bigg( \frac{16}{3} \bigg)
+ \frac{1}{\ep} \bigg( 16
+\frac{3}{2} \zeta_2
\bigg)
+  \bigg(  
20
+24 \zeta_2
-3 \zeta_3
\bigg)
+ \ep \bigg(  
-\frac{364}{3}
+192 \zeta_2
+\frac{489}{40} \zeta_2^2
\nn\\&
+\frac{184}{3} \zeta_3
-96 \zeta_2 \log (2)
\bigg)
+ \ep^2 \bigg(  
-1244
+512 \lih
+1158 \zeta_2
-\frac{394}{5} \zeta_2^2
+760 \zeta_3
\nn\\&
-\frac{261}{2} \zeta_2 \zeta_3
+\frac{627}{5} \zeta_5
-1008 \zeta_2 \log (2)
+160 \zeta_2 \log ^2(2)
+\frac{64 \log ^4(2)}{3}
\bigg)
+ \ep^3 \bigg(  
-7572
\nn\\&
+5376 \lih
+3072 \lifh
+6018 \zeta_2
+864 \lih \zeta_2
-\frac{6474}{5} \zeta_2^2
-\frac{66329}{560} \zeta_2^3
+5340 \zeta_3
\nn\\&
-588 \zeta_2 \zeta_3
-\frac{659}{2} \zeta_3^2
-\frac{9896}{5} \zeta_5
-6768 \zeta_2 \log (2)
+\frac{3888}{5} \zeta_2^2 \log (2)
+756 \zeta_2 \zeta_3 \log (2)
\nn\\&
+1680 \zeta_2 \log ^2(2)
-216 \zeta_2^2 \log ^2(2)
-320 \zeta_2 \log ^3(2)
+224 \log ^4(2)
+36 \zeta_2 \log ^4(2)
\nn\\&
-\frac{128}{5} \log ^5(2)
\bigg)
+ {\mathcal{O}} (\ep^4) \,.
%
\\
 C_{4;2} =&
 \frac{2}{\ep^3} 
+ \frac{1}{\ep^2} \bigg( \frac{23}{3} \bigg)
+ \frac{1}{\ep} \bigg(  
\frac{35}{2}
+3 \zeta_2
\bigg)
+  \bigg(  
\frac{275}{12}
+\frac{23}{2} \zeta_2
-2 \zeta_3
\bigg)
+ \ep \bigg(  
-\frac{189}{8}
+\frac{105}{4} \zeta_2
+\frac{57}{20} \zeta_2^2
\nn\\&
+\frac{89}{3} \zeta_3
\bigg)
+ \ep^2 \bigg(  
-\frac{14917}{48}
+256 \lih
+\frac{275}{8} \zeta_2
-\frac{783}{8} \zeta_2^2
+\frac{525}{2} \zeta_3
-3 \zeta_2 \zeta_3
-\frac{6}{5} \zeta_5
\nn\\&
-64 \zeta_2 \log ^2(2)
+\frac{32 \log ^4(2)}{3}
\bigg)
+ \ep^3 \bigg(  
-\frac{48005}{32}
+1920 \lih
+1536 \lifh
-\frac{567}{16} \zeta_2
\nn\\&
-\frac{12657}{16} \zeta_2^2
+\frac{631}{280} \zeta_2^3
+\frac{15965}{12} \zeta_3
+\frac{89}{2} \zeta_2 \zeta_3
-\frac{6223}{5} \zeta_5
+\frac{3264}{5} \zeta_2^2 \log (2)
-480 \zeta_2 \log ^2(2)
\nn\\&
+128 \zeta_2 \log ^3(2)
+80 \log ^4(2)
-\frac{64}{5} \log ^5(2)
+\zeta_3^2
\bigg)
+ {\mathcal{O}} (\ep^4) \,.
%
%
\\
 C_{4;3} =&
 \frac{1}{\ep^3} \bigg( \frac{3}{2} \bigg)
+ \frac{1}{\ep^2} \bigg( \frac{23}{4} \bigg)
+ \frac{1}{\ep} \bigg(  
\frac{105}{8}
+\frac{9}{4} \zeta_2
\bigg)
+  \bigg(  
\frac{275}{16}
+\frac{133}{8} \zeta_2
-\frac{3}{2} \zeta_3
\bigg)
+ \ep \bigg(  
-\frac{567}{32}
+\frac{1275}{16} \zeta_2
\nn\\&
+\frac{171}{80} \zeta_2^2
+\frac{89}{4} \zeta_3
-48 \zeta_2 \log (2)
\bigg)
+ \ep^2 \bigg(  
-\frac{14917}{64}
+192 \lih
+\frac{10105}{32} \zeta_2
-\frac{941}{32} \zeta_2^2
\nn\\&
+\frac{1575}{8} \zeta_3
-\frac{9}{4} \zeta_2 \zeta_3
-\frac{9}{10} \zeta_5
-360 \zeta_2 \log (2)
+96 \zeta_2 \log ^2(2)
+8 \log ^4(2)
\bigg)
+ \ep^3 \bigg(  
-\frac{144015}{128}
\nn\\&
+1440 \lih
+1152 \lifh
+\frac{72219}{64} \zeta_2
-\frac{16851}{64} \zeta_2^2
+\frac{1893 \zeta_2^3}{1120}
+\frac{15965}{16} \zeta_3
-\frac{437}{8} \zeta_2 \zeta_3
\nn\\&
+\frac{3}{4} \zeta_3^2
-\frac{18669}{20} \zeta_5
-1740 \zeta_2 \log (2)
+\frac{1128}{5} \zeta_2^2 \log (2)
+720 \zeta_2 \log ^2(2)
-192 \zeta_2 \log ^3(2)
\nn\\&
+60 \log ^4(2)
-\frac{48}{5} \log ^5(2)
\bigg)
+ {\mathcal{O}} (\ep^4) \,.
\end{align}
In several intermediate steps of the calculation, we have extensively used 
\textsc{HarmonicSums} \cite{Ablinger:2010kw,Ablinger:2011te,Ablinger:2014rba}, 
and 
\textsc{PolyLogTools} \cite{Duhr:2019tlz}.
We also have evaluated numerically all the appearing MIs using \textsc{AMFlow} \cite{Liu:2022chg} for several 
values of $x$, finding perfect agreement with the numerical evaluation our analytic results. 
Some of the MIs have also been checked numerically against \textsc{FIESTA} \cite{Smirnov:2008py, Smirnov:2021rhf}.

\section{Results} 
\label{sec:results}
In this paper, we present the complete analytic light-fermionic contributions and contributions from 
double heavy quark loops to the three-loop HLFF. 
We present here the hard finite remainder of the HLFF $G_i^{(3)}$s and $S^{(3)}$,
after subtracting the universal IR divergences through eq.~\ref{eq:ir}. 
The finite remainders can also be expanded in $\alpha_s$ as follows
\begin{equation}
 G_i^{\rm fin} = \sum_{n=0}^\infty \asr^n {\mathcal G}_i^{(n)}  \,,
 \quad
 S^{\rm fin} = \sum_{n=0}^\infty \asr^n {\mathcal S}^{(n)}  \,.
\end{equation}
Along with the \texttt{arXiv} submission of this manuscript, 
we provide an ancillary file \texttt{result.nb}
containing all the results of the form factors in \textsc{Mathematica} format.
We choose the renormalization scale to be the top quark mass i.e. $\mu_R^2 = m_t^2$.
For a compact presentation,
we abbreviate some rational functions and a constant that appear frequently in the expressions, as follows:
\begin{align}
 \bar{x} = \frac{1}{1-x} \,, ~~~
 f_n = \frac{1}{x} - n \,,  ~~~ f_{m,n} = \frac{m}{x} - n \,,
 ~~~
 a_4 = \lih + \frac{1}{24} \ln^4 (2) \,.
\end{align}
In the following, we present the relevant results for ${\mathcal{G}_1^{(3)}}$, ${\mathcal{G}_2^{(3)}}$, 
${\mathcal{G}_3^{(3)}}$ and ${\mathcal{S}^{(3)}}$:
\begin{align}
 {\mathcal{G}_1^{(3)}} &= C_F n_l^2 T_F^2  \bigg[
-\frac{322979}{6561}
+\frac{8}{729} (-4919+1827 \ixo) H_1(x)
+\frac{16}{81} (-203+57 \ixo) H_{0,1}(x)
\nn\\
&
+\frac{32}{81} (-203+57 \ixo) H_{1,1}(x)
+\frac{32}{27} (-19+3 \ixo) H_{0,0,1}(x)
+\frac{64}{27} (-19+3 \ixo) H_{0,1,1}(x)
\nn\\
&
+\frac{64}{27} (-19+3 \ixo) H_{1,0,1}(x)
+\frac{128}{27} (-19+3 \ixo) H_{1,1,1}(x)
-\frac{64}{9} H_{0,0,0,1}(x)
-\frac{128}{9} H_{0,0,1,1}(x)
\nn\\
&
-\frac{128}{9} H_{0,1,0,1}(x)
-\frac{256}{9} H_{0,1,1,1}(x)
-\frac{128}{9} H_{1,0,0,1}(x)
-\frac{256}{9} H_{1,0,1,1}(x)
-\frac{256}{9} H_{1,1,0,1}(x)
\nn\\
&
-\frac{512}{9} H_{1,1,1,1}(x)
+\Big(
        -\frac{5384}{81}
        +\frac{64}{27} (-19+3 \ixo) H_1(x)
        -\frac{128}{9} H_{0,1}(x)
        -\frac{256}{9} H_{1,1}(x)
\Big) \zeta_2
\nn\\
&
-\frac{1768}{135} \zeta_2^2
-\frac{64}{243} \zeta_3 \big(
        130+63 H_1(x)\big)
 \bigg]
\nn\\
&
+ C_F^2 n_l T_F  \bigg[
\frac{98977}{972}
+\fxmo \iomx^2 \Big(
        \big(
                1-4 x+5 x^2
        \big)
\Big(-\frac{64}{3} H_{1,-1,0,1}(x)
                -\frac{80}{3} H_{-1,0,0,1}(x)
\nn\\
&
                -32 H_{-1,0,1,1}(x)
                -\frac{64}{3} H_{-1,1,0,1}(x)
                +\frac{64}{3} H_{-1,-1,0,1}(x)
                +\Big(
                        32 \ln (2) H_{-1}(x)
                        -\frac{32}{3} H_{1,-1}(x)
\nn\\
&
                        +\frac{64}{3} H_{-1,1}(x)
                        +\frac{32}{3} H_{-1,-1}(x)
                \Big) \zeta_2
                -8 H_{-1}(x) \zeta_3
        \Big)
        -\frac{16}{9} \big(
                47-146 x+169 x^2\big) H_{-1,0,1}(x)
\nn\\
&
        -\frac{8}{9} \big(
                47-146 x+169 x^2\big) H_{-1}(x) \zeta_2
\Big)
+\iomx \Big(
        \frac{1}{81} \big(
                -23681
                -4116 \ixo
                +216 \ixtw
\nn\\
&
                -7915 x
        \big) H_{1,1}(x)
        +\frac{32}{9} (-110
        +12 \ixo
        +17 x
        ) H_{1,1,1}(x)
\Big)
+\iomx^2 \Big(
        \frac{1}{162} \big(
                -11849
                +1368 \ixo
\nn\\
&
                +44290 x
                +37183 x^2
        \big) H_{0,1}(x)
        -\frac{4}{27} \big(
                -1111
                +153 \ixo
                -1948 x
                -178 x^2
        \big) H_{1,0,1}(x)
\nn\\
&
        +\frac{8}{9} \big(
                 20
                -24 \ixo
                -319 x
                +167 x^2
        \big) H_{1,0,0,1}(x)
        -\frac{16}{9} \big(
                 199
                -15 \ixo
                -173 x
                +133 x^2
        \big) H_{1,0,1,1}(x)
\nn\\
&
        +\frac{16}{9} \big(
                16
                +6 \ixo
                +205 x
                -65 x^2
        \big) H_{1,1,0,1}(x)
        +\Big(
                \frac{1}{648} \big(
                        354679+323122 x-14825 x^2\big)
\nn\\
&
                +32 \ln (2) \big(
                        -1+5 x^2\big) H_1(x)
                -\frac{2}{27} \big(
                        -9779
                        +942 \ixo
                        -7838 x
                        +2107 x^2
                \big) H_1(x)
\nn\\
&
                +\frac{4}{9} \big(
                         778
                        -45 \ixo
                        +1207 x
                        -284 x^2
                \big) H_{1,1}(x)
        \Big) \zeta_2
        -\frac{2}{9} \big(
                 57
                -73 \ixo
                -555 x
                +475 x^2
        \big) H_1(x) \zeta_3
\Big)
\nn\\
&
+\iomx^3 \Big(
        \frac{32}{3} a_4 \big(
                22-42 x+9 x^2+8 x^3\big)
        +\big(
                -1+5 x-3 x^2+x^3
        \big)
\Big(-\frac{128}{3} H_{0,0,-1,0,1}(x)
\nn\\
&
                -\frac{128}{3} H_{0,1,-1,0,1}(x)
                -\frac{160}{3} H_{0,-1,0,0,1}(x)
                -64 H_{0,-1,0,1,1}(x)
                -\frac{128}{3} H_{0,-1,1,0,1}(x)
\nn\\
&
                +\frac{128}{3} H_{0,-1,-1,0,1}(x)
                +\Big(
                        64 \ln (2) H_{0,-1}(x)
                        -\frac{64}{3} H_{0,0,-1}(x)
                        -\frac{64}{3} H_{0,1,-1}(x)
                        +\frac{128}{3} H_{0,-1,1}(x)
\nn\\
&
                        +\frac{64}{3} H_{0,-1,-1}(x)
                \Big) \zeta_2
                -16 H_{0,-1}(x) \zeta_3
        \Big)
        -\frac{8}{9} \big(
                197
                +13 \ixo
                +3 x
                -415 x^2
                +223 x^3
        \big) H_{0,0,1}(x)
\nn\\
&
        +\frac{8}{27} \big(
                87 \ixo
                -4 \big(
                        545-516 x-24 x^2+x^3\big)
        \big) H_{0,1,1}(x)
        +\frac{8}{9} \big(
                28-98 x+64 x^2-15 x^3\big) H_{2,1,1}(x)
\nn\\
&
        -
        \frac{16}{9} \big(
                43
                +6 \ixo
                +94 x
                -250 x^2
                +101 x^3
        \big) H_{0,0,0,1}(x)
        +\frac{16}{9} \big(
                -237
                +6 \ixo
                -57 x
                +56 x^2
\nn\\
&
                +15 x^3
        \big) H_{0,0,1,1}(x)
        -\frac{8}{9} \big(
                -58
                +18 \ixo
                -141 x
                -4 x^2
                -5 x^3
        \big) H_{0,1,0,1}(x)
        +\frac{16}{3} \big(
                -114
                +7 \ixo
\nn\\
&
                +176 x
                -140 x^2
                +77 x^3
        \big) H_{0,1,1,1}(x)
        -\frac{32}{9} \big(
                -53
                +6 \ixo
                +151 x
                -45 x^2
                -13 x^3
        \big) H_{0,-1,0,1}(x)
\nn\\
&
        +\frac{32}{3} \big(
                -2-2 x+x^2\big) H_{2,1,1,1}(x)
        -\frac{16}{3} \big(
                -2-2 x+x^2\big) H_{2,2,1,1}(x)
\nn\\
&
        -\frac{64}{3} \big(
                1+10 x-4 x^2+2 x^3\big) H_{0,0,0,0,1}(x)
        -\frac{64}{3} \big(
                5+14 x+x^2+x^3\big) H_{0,0,0,1,1}(x)
\nn\\
&
        -\frac{16}{3} \big(
                -4+20 x-19 x^2+7 x^3\big) H_{0,0,1,0,1}(x)
        +\frac{32}{3} \big(
                -15-31 x-17 x^2+3 x^3\big) H_{0,0,1,1,1}(x)
\nn\\
&
        -\frac{16}{3} \big(
                -2+44 x-27 x^2+11 x^3\big) H_{0,1,0,0,1}(x)
        +\frac{32}{3} \big(
                -9-5 x-13 x^2+3 x^3\big) H_{0,1,0,1,1}(x)
\nn\\
&
        +\frac{32}{3} \big(
                2+29 x-8 x^2+4 x^3\big) H_{0,1,1,0,1}(x)
        +\Big(
                \frac{32}{3} \ln^2 (2) \big(
                        14-18 x+3 x^2+4 x^3\big)
\nn\\
&
                +\ln (2) \Big(
                        -\frac{8}{3} \big(
                                232-526 x+260 x^2+13 x^3\big)
                        +16 \big(
                                -2-2 x+x^2\big) H_2(x)
                        +128 x H_{0,1}(x)
                \Big)
\nn\\
&
                -
                \frac{8}{3} \big(
                        -28+98 x-64 x^2+15 x^3\big) H_2(x)
                -\frac{2}{9} \big(
                        -1999
                        +201 \ixo
                        -605 x
\nn\\
&
                        +1419 x^2
                        -776 x^3
                \big) H_{0,1}(x)
                -\frac{16}{9} \big(
                        -53
                        +6 \ixo
                        +151 x
                        -45 x^2
                        -13 x^3
                \big) H_{0,-1}(x)
\nn\\
&
                +16 \big(
                        -2-2 x+x^2\big) H_{2,1}(x)
                -16 \big(
                        -2-2 x+x^2\big) H_{2,2}(x)
\nn\\
&
                -\frac{4}{3} \big(
                        -107+81 x-181 x^2+47 x^3\big) H_{0,0,1}(x)
                +8 \big(
                        21+61 x+7 x^2+3 x^3\big) H_{0,1,1}(x)
\nn\\
&
                +\Big(
                        -\frac{725}{9}-1225 x+393 x^2-\frac{1723 x^3}{9} \Big) \zeta_3
        \Big) \zeta_2
        +\frac{1}{270} \big(
                -48931-52407 x+59343 x^2
\nn\\
&                
                -44381 x^3\big) \zeta_2^2
        +\Big(
                \frac{2}{81} \big(
                        6826-7032 x-4038 x^2+4811 x^3\big)
                +\frac{28}{3} \big(
                        -2-2 x+x^2\big) H_2(x)
\nn\\
&
                +\frac{4}{9} \big(
                        -49+243 x-231 x^2+85 x^3\big) H_{0,1}(x)
        \Big) \zeta_3
        +\frac{8}{9} \big(
                104+111 x+144 x^2-32 x^3\big) \zeta_5
\Big)
\nn\\
&
+\frac{1}{324} (-4071+10265 \ixo) H_1(x)
+\frac{320}{3} (-5+\ixo) H_{1,1,1,1}(x)
+\frac{32}{3} \ixo H_{1,2,1,1}(x)
\nn\\
&
-\frac{640}{3} H_{0,1,1,1,1}(x)
-\frac{64}{3} H_{0,1,2,1,1}(x)
+128 H_{1,0,0,0,1}(x)
+\frac{128}{3} H_{1,0,1,0,1}(x)
\nn\\
&
-192 H_{1,0,1,1,1}(x)
+\frac{352}{3} H_{1,1,0,0,1}(x)
-128 H_{1,1,0,1,1}(x)
-\frac{256}{3} H_{1,1,1,0,1}(x)
\nn\\
&
-\frac{1280}{3} H_{1,1,1,1,1}(x)
-\frac{64}{3} H_{1,1,2,1,1}(x)
+\Big(
        64 \ln (2) H_{1,1}(x)
        +32 \ixo H_{1,2}(x)
        -64 H_{0,1,2}(x)
\nn\\
&
        -\frac{8}{3} H_{1,0,1}(x)
        -\frac{208}{3} H_{1,1,1}(x)
        -64 H_{1,1,2}(x)
\Big) \zeta_2
+\frac{488}{15} H_1(x) \zeta_2^2
-\frac{728}{9} H_{1,1}(x) \zeta_3
\bigg]
\nn\\
&
+ C_F C_A n_l T_F   \bigg[ 
\frac{2866346}{6561}
+\fxmo \iomx^2 \Big(
        \big(
                1-4 x+5 x^2
        \big)
\Big(\frac{32}{3} H_{1,-1,0,1}(x)
                +\frac{40}{3} H_{-1,0,0,1}(x)
\nn\\
&
                +16 H_{-1,0,1,1}(x)
                +\frac{32}{3} H_{-1,1,0,1}(x)
                -\frac{32}{3} H_{-1,-1,0,1}(x)
                +\Big(
                        -16 \ln (2) H_{-1}(x)
                        +\frac{16}{3} H_{1,-1}(x)
\nn\\
&
                        -\frac{32}{3} H_{-1,1}(x)
                        -\frac{16}{3} H_{-1,-1}(x)
                \Big) \zeta_2
                +4 H_{-1}(x) \zeta_3
        \Big)
        +\frac{8}{9} \big(
                47-146 x+169 x^2\big) H_{-1,0,1}(x)
\nn\\
&
        +\frac{4}{9} \big(
                47-146 x+169 x^2\big) H_{-1}(x) \zeta_2
\Big)
+\iomx \Big(
        -\frac{2}{81} (-19682
        +5061 \ixo
        +20813 x
        ) H_{1,1}(x)
\nn\\
&
        -\frac{4}{27} (-3074
        +393 \ixo
        +3329 x
        ) H_{1,1,1}(x)
\Big)
+\iomx^2 \Big(
        -\frac{2}{81} \big(
                -14725
                +3138 \ixo
\nn\\
&
                +14456 x
                -9061 x^2
        \big) H_{0,1}(x)
        -\frac{2}{27} \big(
                -5021
                +501 \ixo
                +6181 x
                -3599 x^2
        \big) H_{1,0,1}(x)
\nn\\
&
        +\frac{8}{3} \big(
                -5
                +2 \ixo
                -26 x
                -8 x^2
        \big) H_{1,0,0,1}(x)
        +\frac{16}{9} \big(
                58-203 x+91 x^2\big) H_{1,0,1,1}(x)
\nn\\
&
        +\frac{4}{9} \big(
                397-569 x+370 x^2\big) H_{1,1,0,1}(x)
        +\Big(
                \frac{1}{729} \big(
                        262867-137744 x+292189 x^2\big)
\nn\\
&
                -16 \ln (2)
                 \big(
                        -1+5 x^2\big) H_1(x)
                +
                \frac{4}{81} \big(
                        8257
                        +54 \ixo
                        -3851 x
                        +6565 x^2
                \big) H_1(x)
\nn\\
&
                +\frac{4}{9} \big(
                        323
                        +48 \ixo
                        -7 x
                        +302 x^2
                \big) H_{1,1}(x)
        \Big) \zeta_2
        -\frac{4}{3} \big(
                -35
                +6 \ixo
                +30 x
                -47 x^2
        \big) H_1(x) \zeta_3
\Big)
\nn\\
&
+\iomx^3 \Big(
        -\frac{16}{3} a_4 \big(
                22-42 x+9 x^2+8 x^3\big)
\nn\\
&
        +\big(
                -1+5 x-3 x^2+x^3
        \big)
\Big( \frac{64}{3} H_{0,0,-1,0,1}(x)
                +\frac{64}{3} H_{0,1,-1,0,1}(x)
                +\frac{80}{3} H_{0,-1,0,0,1}(x)
\nn\\
&
                +32 H_{0,-1,0,1,1}(x)
                +\frac{64}{3} H_{0,-1,1,0,1}(x)
                -\frac{64}{3} H_{0,-1,-1,0,1}(x)
                +\Big(
                        -32 \ln (2) H_{0,-1}(x)
\nn\\
&
                        +\frac{32}{3} H_{0,0,-1}(x)
                        +\frac{32}{3} H_{0,1,-1}(x)
                        -\frac{64}{3} H_{0,-1,1}(x)
                        -\frac{32}{3} H_{0,-1,-1}(x)
                \Big) \zeta_2
                +8 H_{0,-1}(x) \zeta_3
        \Big)
\nn\\
&
        -\frac{2}{27} \big(
                -1006
                +264 \ixo
                +4533 x
                -4350 x^2
                +433 x^3
        \big) H_{0,0,1}(x)
        -\frac{4}{27} \big(
                -1429
                +210 \ixo
\nn\\
&
                +5340 x
                -5799 x^2
                +1741 x^3
        \big) H_{0,1,1}(x)
        +\frac{4}{9} \big(
                -28+98 x-64 x^2+15 x^3\big) H_{2,1,1}(x)
\nn\\
&
        +\frac{4}{9} \big(
                -43-467 x+76 x^2+107 x^3\big) H_{0,0,0,1}(x)
        -\frac{8}{9}
         \big(
                25+307 x-265 x^2+31 x^3\big) H_{0,0,1,1}(x)
\nn\\
&
        -
        \frac{4}{9} \big(
                -142+413 x-497 x^2+101 x^3\big) H_{0,1,0,1}(x)
        -\frac{8}{9} \big(
                -83+603 x-723 x^2
\nn\\
&
                +221 x^3\big) H_{0,1,1,1}(x)
        +\frac{16}{9} \big(
                -53
                +6 \ixo
                +151 x
                -45 x^2
                -13 x^3
        \big) H_{0,-1,0,1}(x)
\nn\\
&
        -\frac{16}{3} \big(
                -2-2 x+x^2\big) H_{2,1,1,1}(x)
        +\frac{8}{3} \big(
                -2-2 x+x^2\big) H_{2,2,1,1}(x)
\nn\\
&
        -\frac{8}{3} \big(
                4+31 x+16 x^2\big) H_{0,0,0,0,1}(x)
        -\frac{16}{3} \big(
                4+25 x+16 x^2\big) H_{0,0,0,1,1}(x)
\nn\\
&
        -\frac{8}{3} \big(
                1+18 x+4 x^2\big) H_{0,0,1,0,1}(x)
        -\frac{32}{3} \big(
                2+5 x+8 x^2\big) H_{0,0,1,1,1}(x)
\nn\\
&
        +\frac{8}{3} \big(
                -7-10 x-24 x^2+4 x^3\big) H_{0,1,0,0,1}(x)
        -\frac{32}{3} (1+2 x)^2 H_{0,1,0,1,1}(x)
\nn\\
&
        +\frac{8}{3} \big(
                4+13 x+16 x^2\big) H_{0,1,1,0,1}(x)
        +\Big(
                -\frac{16}{3} \ln^2 (2) \big(
                        14-18 x+3 x^2+4 x^3\big)
\nn\\
&
                +\ln (2) \Big(
                        \frac{4}{3} \big(
                                232-526 x+260 x^2+13 x^3\big)
                        -8 \big(
                                -2-2 x+x^2\big) H_2(x)
                        -64 x H_{0,1}(x)
                \Big)
\nn\\
&
                +\frac{4}{3} \big(
                        -28+98 x-64 x^2+15 x^3\big) H_2(x)
                +\frac{4}{9} \big(
                        81
                        +24 \ixo
                        +286 x
                        +342 x^2
                        -228 x^3
                \big) H_{0,1}(x)
\nn\\
&
                +\frac{8}{9} \big(
                        -53
                        +6 \ixo
                        +151 x
                        -45 x^2
                        -13 x^3
                \big) H_{0,-1}(x)
                -8 \big(
                        -2-2 x+x^2\big) H_{2,1}(x)
\nn\\
&
                +8 \big(
                        -2-2 x+x^2\big) H_{2,2}(x)
                +\frac{8}{3} \big(
                        -3+52 x-4 x^2+8 x^3\big) H_{0,0,1}(x)
\nn\\
&
                +\frac{8}{3} \big(
                        -4+99 x+16 x^3\big) H_{0,1,1}(x)
                +\frac{4}{3} \big(
                        -53-43 x-193 x^2+19 x^3\big) \zeta_3
        \Big) \zeta_2
\nn\\
&
        +\frac{2}{135} \big(
                626-12429 x-5394 x^2+2479 x^3\big) \zeta_2^2
        +\Big(
                \frac{1}{81} \big(
                        2080-12666 x+22548 x^2-12529 x^3\big)
\nn\\
&
                -\frac{14}{3} \big(
                        -2-2 x+x^2\big) H_2(x)
                -\frac{8}{3} \big(
                        -7-22 x^2+6 x^3\big) H_{0,1}(x)
        \Big) \zeta_3
        +\frac{4}{3} \big(
                -43+157 x
\nn\\
&                
                -121 x^2+51 x^3\big) \zeta_5
\Big)
+\frac{2}{729} (154919-63414 \ixo) H_1(x)
+\frac{2816}{9} H_{1,1,1,1}(x)
-\frac{16}{3} \ixo H_{1,2,1,1}(x)
\nn\\
&
+\frac{32}{3} H_{0,1,2,1,1}(x)
-\frac{176}{3} H_{1,0,0,0,1}(x)
-\frac{128}{3} H_{1,0,0,1,1}(x)
-\frac{64}{3} H_{1,0,1,0,1}(x)
-64 H_{1,1,0,0,1}(x)
\nn\\
&
-\frac{32}{3} H_{1,1,0,1,1}(x)
+\frac{32}{3} H_{1,1,2,1,1}(x)
+\Big(
        -32 \ln (2) H_{1,1}(x)
        -16 \ixo H_{1,2}(x)
        +32 H_{0,1,2}(x)
\nn\\
&
        -\frac{128}{3} H_{1,1,1}(x)
        +32 H_{1,1,2}(x)
\Big) \zeta_2
-\frac{152}{3} H_1(x) \zeta_2^2
+24 H_{1,1}(x) \zeta_3
\bigg] 
%
%
\nn\\
&
+ C_F n_l n_h T_F^2   \bigg[ 
\fxmo \iomx^3 \big(
        -3-5 x+5 x^2+19 x^3
\big)
\Big(-\frac{64}{27} H_{0,1,1}(x)
        -\frac{64}{27} H_{1,0,1}(x)
        +\frac{64}{27} H_1(x) \zeta_2
\Big)
\nn\\
&
+\iomx^2 \Big(
        -\frac{4}{729} \big(
                37321-74966 x+16909 x^2\big)
        +\ixo \Big(
                \frac{3248}{81} H_1(x)
                +\frac{608}{27} H_{1,1}(x)
        \Big)
        -\frac{16}{81} \big(
                467
\nn\\
&
                -919 x+463 x^2\big) H_1(x)
        -\frac{32}{81} \big(
                -89+73 x+265 x^2\big) H_{1,1}(x)
\Big)
+\iomx^3 \Big(
        \ixo \Big(
                \frac{608}{27} H_{0,1}(x)
\nn\\
&
                +\frac{64}{9} H_{0,0,1}(x)
        \Big)
        +\big(
                -1-3 x-3 x^2+x^3
        \big)
\Big(\frac{128}{9} H_{0,0,0,1}(x)
                +\frac{128}{9} H_{0,0,1,1}(x)
                +\frac{128}{9} H_{0,1,0,1}(x)
\nn\\
&
                -\frac{128}{9} H_{0,1}(x) \zeta_2
                +\frac{128}{45} \zeta_2^2
        \Big)
        +\frac{32}{81} \big(
                -68+9 x+36 x^2+62 x^3\big) H_{0,1}(x)
        -\frac{64}{27} \big(
                10-15 x+60 x^2
\nn\\
&
                +19 x^3\big) H_{0,0,1}(x)
        +\frac{16}{81} \big(
                317-1179 x+399 x^2+271 x^3\big) \zeta_2
        +\frac{64}{9} \big(
                6-5 x+12 x^2\big) \zeta_3
\Big)
\bigg]
%
%
\nn\\
&
+ C_F n_h^2 T_F^2   \bigg[
\iomx^4 \Big(
        -\frac{2}{729} \big(
                20951-117716 x+221898 x^2-182516 x^3+55655 x^4\big)
\nn\\
&
        +\frac{1624}{81} \ixo H_1(x)
        -\frac{8}{243} \big(
                -29+1658 x-54 x^2-3145 x^3+1891 x^4\big) H_1(x)
\Big)
\nn\\
&
+\iomx^6 \Big(
        \frac{32}{9} \ixo H_{0,0,1}(x)
        -\frac{32}{27} \big(
                1+27 x-81 x^2+21 x^3+45 x^4-45 x^5+19 x^6\big) H_{0,0,1}(x)
\nn\\
&
        +\frac{32}{27} \big(
                7-180 x+687 x^2-1206 x^3+1035 x^4-438 x^5+79 x^6\big) \zeta_3
\Big)
+\iomx^3 \Big(
        \frac{64}{9} \big(
                -1-9 x-3 x^2
\nn\\
&
                +x^3\big) H_{0,0,0,1}(x)
        +\frac{128}{9} \big(
                -1-9 x-3 x^2+x^3\big) H_{1,0,0,1}(x)
        +\frac{64}{15} \big(
                -1-9 x-3 x^2+x^3\big) \zeta_2^2
\nn\\
&
        -\frac{128}{9} \big(
                -1-9 x-3 x^2+x^3\big) H_1(x) \zeta_3
\big)
-\frac{16 \fxmo \iomx^5}{81} \big(
        -57-67 x+312 x^2-184 x^3
\nn\\
&
        -255 x^4+203 x^5\big) H_{0,1}(x)
+\frac{128}{45} \iomx (5+3 x) \zeta_2
\bigg] \,.
\end{align}

\begin{align}
 {\mathcal{G}_2^{(3)}} &= C_F n_l^2 T_F^2  \bigg[
 \ixo \big(
        -\frac{5072}{81} H_1(x)
        -\frac{800}{27} H_{0,1}(x)
        -\frac{1600}{27} H_{1,1}(x)
        -\frac{64}{9} H_{0,0,1}(x)
        -\frac{128}{9} H_{0,1,1}(x)
\nn\\
&
        -\frac{128}{9} H_{1,0,1}(x)
        -\frac{256}{9} H_{1,1,1}(x)
        -\frac{128}{9} H_1(x) \zeta_2
\big)   
\bigg] 
%
%
\nn\\
&
+ C_F^2 n_l T_F  \bigg[
\iomx \Big(
        -
        \frac{8}{3}
        +\frac{1}{162} (20201-13865 \ixo) H_1(x)
\Big)
\nn\\
&
+\iomx^2 \Big(
        -\frac{8}{27} \big(
                -6778
                -361 \ixo
                +18 \ixtw
                -1753 x
        \big) H_{1,1}(x)
        -\frac{64}{3} (-78
        +5 \ixo
        -8 x
        ) H_{1,1,1}(x)
\Big)
\nn\\
&
+\fxmo \Big(
        \iomx^3 \Big(
                \big(
                        1+4 x+x^2
                \big)
\Big(\frac{128}{3} H_{1,-1,0,1}(x)
                        +\frac{160}{3} H_{-1,0,0,1}(x)
                        +64 H_{-1,0,1,1}(x)
\nn\\
&
                        +\frac{128}{3} H_{-1,1,0,1}(x)
                        -\frac{128}{3} H_{-1,-1,0,1}(x)
                        +\Big(
                                -64 \ln (2) H_{-1}(x)
                                +\frac{64}{3} H_{1,-1}(x)
                                -\frac{128}{3} H_{-1,1}(x)
\nn\\
&
                                -\frac{64}{3} H_{-1,-1}(x)
                        \Big) \zeta_2
                        +16 H_{-1}(x) \zeta_3
                \Big)
                +\frac{32}{9} \big(
                        59+68 x+83 x^2\big) H_{-1,0,1}(x)
                +\frac{16}{9} \big(
                        59+68 x
\nn\\
&
                        +83 x^2\big) H_{-1}(x) \zeta_2
        \Big)
        -\frac{640}{3} \iomx^4 x (1+2 x) H_{0,0,1}(x) \zeta_2
\Big)
+\iomx^3 \Big(
        \frac{8}{9} \big(
                239
                +7 \ixo
                -2626 x
\nn\\
&
                -578 x^2
        \big) H_{0,1}(x)
        +\frac{8}{9} \big(
                -741
                +81 \ixo
                -2348 x
                -76 x^2
        \big) H_{1,0,1}(x)
        +\frac{16}{3} \big(
                62
                +8 \ixo
                +107 x
\nn\\
&
                -5 x^2
        \big) H_{1,0,0,1}(x)
        -\frac{32}{3} \big(
                -83
                +5 \ixo
                -49 x
                -17 x^2
        \big) H_{1,0,1,1}(x)
        -
        \frac{32}{3} \big(
                53
                +2 \ixo
                +115 x
\nn\\
&
                -8 x^2
        \big) H_{1,1,0,1}(x)
        +\big(
                -\frac{8}{3} \big(
                        8+2235 x+59 x^2\big)
                -128 \ln (2) \big(
                        4+x+x^2\big) H_1(x)
\nn\\
&
                -\frac{4}{9} \big(
                         6109
                        -491 \ixo
                        +8281 x
                        +669 x^2
                \big) H_1(x)
                +8 \big(
                        -235
                        +5 \ixo
                        -309 x
                        -13 x^2
                \big) H_{1,1}(x)
        \big) \zeta_2
\nn\\
&
        -\frac{4}{9} \big(
                93
                +73 \ixo
                +471 x
                -157 x^2
        \big) H_1(x) \zeta_3
\Big)
+\iomx^4 \big(
        -\frac{64}{3} a_4 \big(
                -30+54 x-50 x^2+17 x^3\big)
\nn\\
&
        +\frac{16}{9} \big(
                328
                +25 \ixo
                -134 x
                -239 x^2
                +83 x^3
        \big) H_{0,0,1}(x)
        -\frac{16}{9} \big(
                -1380
                +29 \ixo
                +700 x
\nn\\
&
                +715 x^2
                -x^3
        \big) H_{0,1,1}(x)
        -\frac{16}{9} \big(
                -104+90 x-71 x^2+22 x^3\big) H_{2,1,1}(x)
        +\frac{32}{9} \big(
                121
                +6 \ixo
\nn\\
&
                -18 x
                -127 x^2
        \big) H_{0,0,0,1}(x)
        -\frac{32}{9} \big(
                -368
                +6 \ixo
                -564 x
                +260 x^2
                +15 x^3
        \big) H_{0,0,1,1}(x)
\nn\\
&
        +\frac{16}{9} \big(
                10
                +18 \ixo
                -531 x
                -40 x^2
                -27 x^3
        \big) H_{0,1,0,1}(x)
        -\frac{32}{3} \big(
                -160
                +7 \ixo
                +2 x
                +152 x^2
\nn\\
&
                +17 x^3
        \big) H_{0,1,1,1}(x)
        +\frac{128}{3} \big(
                -1
                +\ixo
                +31 x
                -9 x^2
                +x^3
        \big) H_{0,-1,0,1}(x)
\nn\\
&
        -
        \frac{64}{3} \big(
                -14+6 x-2 x^2+x^3\big) H_{2,1,1,1}(x)
        +\frac{32}{3} \big(
                -14+6 x-2 x^2+x^3\big) H_{2,2,1,1}(x)
\nn\\
&
        +\frac{128}{3} \big(
                4+15 x+8 x^2\big) H_{0,0,0,0,1}(x)
        +\frac{128}{3} \big(
                8+39 x+16 x^2\big) H_{0,0,0,1,1}(x)
\nn\\
&
        +\frac{128}{3} \big(
                1+2 x^2\big) H_{0,0,1,0,1}(x)
        +\frac{256}{3} (4+x) (1+8 x) H_{0,0,1,1,1}(x)
        +512 x H_{0,0,-1,0,1}(x)
\nn\\
&
        +64 \big(
                2+7 x+4 x^2\big) H_{0,1,0,0,1}(x)
        +\frac{512}{3} \big(
                1+6 x+2 x^2\big) H_{0,1,0,1,1}(x)
        -\frac{64}{3} \big(
                8+57 x
\nn\\
&
                +16 x^2\big) H_{0,1,1,0,1}(x)
        +512 x H_{0,1,-1,0,1}(x)
        +640 x H_{0,-1,0,0,1}(x)
        +768 x H_{0,-1,0,1,1}(x)
\nn\\
&
        +512 x H_{0,-1,1,0,1}(x)
        -512 x H_{0,-1,-1,0,1}(x)
        +\big(
                -\frac{64}{3} \ln^2 (2) \big(
                        6+18 x-22 x^2+7 x^3\big)
\nn\\
&
                +\ln (2) \big(
                        \frac{16}{3} \big(
                                -352+618 x-527 x^2+198 x^3\big)
                        -32 \big(
                                -14+6 x-2 x^2+x^3\big) H_2(x)
\nn\\
&
                        -768 x H_{0,1}(x)
                        -768 x H_{0,-1}(x)
                \big)
                -\frac{16}{3} \big(
                        -104+90 x-71 x^2+22 x^3\big) H_2(x)
\nn\\
&
                +\frac{4}{3} \big(
                        -692
                        +67 \ixo
                        -1338 x
                        +196 x^2
                        +7 x^3
                \big) H_{0,1}(x)
                +\frac{64}{3} \big(
                        -1
                        +\ixo
                        +31 x
                        -9 x^2
\nn\\
&
                        +x^3
                \big) H_{0,-1}(x)
                -32 \big(
                        -14+6 x-2 x^2+x^3\big) H_{2,1}(x)
                +32 \big(
                        -14+6 x-2 x^2+x^3\big) H_{2,2}(x)
\nn\\
&
                +256 x H_{0,0,-1}(x)
                -64 \big(
                        8+45 x+16 x^2\big) H_{0,1,1}(x)
                +256 x H_{0,1,-1}(x)
                -512 x H_{0,-1,1}(x)
\nn\\
&
                -256 x H_{0,-1,-1}(x)
                +\frac{32}{3} \big(
                        68+417 x+136 x^2\big) \zeta_3
        \big) \zeta_2
        -\frac{8}{15} \big(
                448-4168 x+516 x^2-395 x^3\big) \zeta_2^2
\nn\\
&
        +\big(
                -\frac{4}{9} \big(
                        -1048+3246 x-3115 x^2+1106 x^3\big)
                -\frac{56}{3} \big(
                        -14+6 x-2 x^2+x^3\big) H_2(x)
\nn\\
&
                -\frac{128}{3} \big(
                        1+2 x^2\big) H_{0,1}(x)
                +192 x H_{0,-1}(x)
        \big) \zeta_3
        -\frac{16}{3} \big(
                32+231 x+64 x^2\big) \zeta_5
\big)
\nn\\
&
-\frac{640}{3} \ixo H_{1,1,1,1}(x)
-\frac{64}{3} \ixo H_{1,2,1,1}(x)
-64 \ixo H_{1,2}(x) \zeta_2
\bigg] 
%
%
\nn\\
&
+ C_F C_A n_l T_F    \bigg[ 
\fxmo \iomx^3 \Big(
        \big(
                1+4 x+x^2
        \big)
\Big(-
                \frac{64}{3} H_{1,-1,0,1}(x)
                -\frac{80}{3} H_{-1,0,0,1}(x)
                -32 H_{-1,0,1,1}(x)
\nn\\
&
                -\frac{64}{3} H_{-1,1,0,1}(x)
                +\frac{64}{3} H_{-1,-1,0,1}(x)
                +\Big(
                        32 \ln (2) H_{-1}(x)
                        -\frac{32}{3} H_{1,-1}(x)
                        +\frac{64}{3} H_{-1,1}(x)
\nn\\
&
                        +\frac{32}{3} H_{-1,-1}(x)
                \Big) \zeta_2
                -8 H_{-1}(x) \zeta_3
        \Big)
        -\frac{16}{9} \big(
                59+68 x+83 x^2\big) H_{-1,0,1}(x)
        -\frac{8}{9} \big(
                59+68 x
\nn\\
&
                +83 x^2\big) H_{-1}(x) \zeta_2
\Big)
+\iomx \Big(
        \frac{248}{3}
        +\frac{16}{81} (-1385+2546 \ixo) H_1(x)
\Big)
+\iomx^2 \Big(
        \frac{4}{27} (682
        +2107 \ixo
\nn\\
&
        +3403 x
        ) H_{1,1}(x)
        +\frac{8}{9} (206
        +131 \ixo
        +311 x
        ) H_{1,1,1}(x)
\Big)
+\iomx^3 \Big(
        \frac{4}{27} \big(
                -387
                +1310 \ixo
\nn\\
&
                -4026 x
                -3089 x^2
        \big) H_{0,1}(x)
        +\frac{4}{9} \big(
                -581
                +167 \ixo
                -871 x
                -653 x^2
        \big) H_{1,0,1}(x)
        +\frac{16}{3} \big(
                 23
                -2 \ixo
\nn\\
&
                +74 x
                +8 x^2
        \big) H_{1,0,0,1}(x)
        -\frac{32}{3} \big(
                -6-49 x+x^2\big) H_{1,0,1,1}(x)
        -\frac{8}{3} \big(
                48+121 x+29 x^2\big) H_{1,1,0,1}(x)
\nn\\
&
        +\big(
                -\frac{4}{9} \big(
                        1786+4239 x+1703 x^2\big)
                +64 \ln (2)
                 \big(
                        4+x+x^2\big) H_1(x)
                -
                \frac{8}{9} \big(
                        800
                        +30 \ixo
                        +2522 x
\nn\\
&
                        +323 x^2
                \big) H_1(x)
                -\frac{8}{3} \big(
                        96
                        +16 \ixo
                        +519 x
                        +35 x^2
                \big) H_{1,1}(x)
        \big) \zeta_2
        +\frac{16}{3} \big(
                -22
                +3 \ixo
                -36 x
\nn\\
&
                -6 x^2
        \big) H_1(x) \zeta_3
\Big)
+\iomx^4 \big(
        \frac{32}{3} a_4 \big(
                -30+54 x-50 x^2+17 x^3\big)
        +\frac{4}{9} \big(
                858
                +88 \ixo
                +1151 x
\nn\\
&
                -2224 x^2
                +x^3
        \big) H_{0,0,1}(x)
        +\frac{8}{9} \big(
                402
                +70 \ixo
                +581 x
                -1155 x^2
                +165 x^3
        \big) H_{0,1,1}(x)
\nn\\
&
        +\frac{8}{9} \big(
                -104+90 x-71 x^2+22 x^3\big) H_{2,1,1}(x)
        -\frac{8}{9} \big(
                -265-1217 x+459 x^2+42 x^3\big) H_{0,0,0,1}(x)
\nn\\
&
        -\frac{16}{9} \big(
                -211-509 x+381 x^2+45 x^3\big) H_{0,0,1,1}(x)
        +\frac{8}{9} \big(
                53+118 x-561 x^2+15 x^3\big) H_{0,1,0,1}(x)
\nn\\
&
        -\frac{32}{3} \big(
                -23-62 x+71 x^2+5 x^3\big) H_{0,1,1,1}(x)
        -\frac{64}{3} \big(
                -1
                +\ixo
                +31 x
                -9 x^2
                +x^3
        \big) H_{0,-1,0,1}(x)
\nn\\
&
        +\frac{32}{3} \big(
                -14+6 x-2 x^2+x^3\big) H_{2,1,1,1}(x)
        -\frac{16}{3} \big(
                -14+6 x-2 x^2+x^3\big) H_{2,2,1,1}(x)
\nn\\
&
        +\frac{16}{3} \big(
                8+97 x+48 x^2\big) H_{0,0,0,0,1}(x)
        +\frac{32}{3} \big(
                8+79 x+48 x^2\big) H_{0,0,0,1,1}(x)
        +
        \frac{16}{3} \big(
                2+55 x
\nn\\
&
                +12 x^2\big) H_{0,0,1,0,1}(x)
        +\frac{64}{3} \big(
                4+17 x+24 x^2\big) H_{0,0,1,1,1}(x)
        -256 x H_{0,0,-1,0,1}(x)
        +16 \big(
                2+23 x
\nn\\
&
                +12 x^2\big) H_{0,1,0,0,1}(x)
        +\frac{64}{3} \big(
                2+13 x+12 x^2\big) H_{0,1,0,1,1}(x)
        -\frac{16}{3} \big(
                8+43 x+48 x^2\big) H_{0,1,1,0,1}(x)
\nn\\
&
        -256 x H_{0,1,-1,0,1}(x)
        -320 x H_{0,-1,0,0,1}(x)
        -384 x H_{0,-1,0,1,1}(x)
        -256 x H_{0,-1,1,0,1}(x)
\nn\\
&
        +256 x H_{0,-1,-1,0,1}(x)
        +\Big(
                \frac{32}{3} \ln^2 (2) \big(
                        6+18 x-22 x^2+7 x^3\big)
                +\ln (2) \big(
                        -\frac{8}{3} \big(
                                -352+618 x
\nn\\
&
                                -527 x^2+198 x^3\big)
                        +16 \big(
                                -14+6 x-2 x^2+x^3\big) H_2(x)
                        +384 x H_{0,1}(x)
                        +384 x H_{0,-1}(x)
                \big)
\nn\\
&
                +\frac{8}{3} \big(
                        -104+90 x-71 x^2+22 x^3\big) H_2(x)
                -\frac{8}{3} \big(
                        77
                        +8 \ixo
                        +426 x
                        +23 x^2
                        -29 x^3
                \big) H_{0,1}(x)
\nn\\
&
                -\frac{32}{3} \big(
                        -1
                        +\ixo
                        +31 x
                        -9 x^2
                        +x^3
                \big) H_{0,-1}(x)
                +16 \big(
                        -14+6 x-2 x^2+x^3\big) H_{2,1}(x)
\nn\\
&
                -16 \big(
                        -14+6 x-2 x^2+x^3\big) H_{2,2}(x)
                -\frac{16}{3} \big(
                        10+89 x+60 x^2\big) H_{0,0,1}(x)
                -128 x H_{0,0,-1}(x)
\nn\\
&
                -16 \big(
                        8+55 x+48 x^2\big) H_{0,1,1}(x)
                -128 x H_{0,1,-1}(x)
                +256 x H_{0,-1,1}(x)
                +128 x H_{0,-1,-1}(x)
\nn\\
&
                +\frac{16}{3} \big(
                        34+167 x+204 x^2\big) \zeta_3
        \Big) \zeta_2
        -\frac{4}{15} \big(
                -697-1934 x-2433 x^2+158 x^3\big) \zeta_2^2
\nn\\
&
        +\Big(
                \frac{2}{9} \big(
                        -1312+1962 x-1699 x^2+1238 x^3\big)
                +\frac{28}{3} \big(
                        -14+6 x-2 x^2+x^3\big) H_2(x)
\nn\\
&
                -\frac{16}{3} \big(
                        2+55 x+12 x^2\big) H_{0,1}(x)
                -96 x H_{0,-1}(x)
        \Big) \zeta_3
        -256 x^2 \zeta_5
        -\frac{32}{3} (4+5 x) \zeta_5
\big)
\nn\\
&
+\frac{32}{3} \ixo H_{1,2,1,1}(x)
+32 \ixo H_{1,2}(x) \zeta_2
\bigg]
%
%
\nn\\
&
+ C_F n_l n_h T_F^2    \bigg[ 
\fxmo \iomx^3 \Big(
        -\frac{128}{9} \big(
                1+20 x+x^2\big) H_{0,1,1}(x)
        -\frac{128}{9} \big(
                1+20 x+x^2\big) H_{1,0,1}(x)
\nn\\
&
        +\frac{128}{9} \big(
                1+20 x+x^2\big) H_1(x) \zeta_2
\Big)
+\iomx^2 \Big(
         \frac{32}{27} (282-35 x)
        -\frac{32}{81} (668
        +317 \ixo
        +317 x
        ) H_1(x)
\nn\\
&
        -\frac{64}{27} (322
        +25 \ixo
        +25 x
        ) H_{1,1}(x)
\Big)
+\iomx^3 \Big(
        -\frac{64}{27} (159
        +25 \ixo
        -138 x
        ) H_{0,1}(x)
\nn\\
&
        +\frac{128}{27} \big(
                6+105 x+29 x^2\big) \zeta_2
\Big)
+\iomx^4 \Big(
        -\frac{128}{9} \big(
                8
                +\ixo
                -15 x
                -32 x^2
                -x^3
        \big) H_{0,0,1}(x)
\nn\\
&
        +512 x H_{0,0,0,1}(x)
        +512 x H_{0,0,1,1}(x)
        +512 x H_{0,1,0,1}(x)
        -512 x H_{0,1}(x) \zeta_2
        +\frac{512}{5} x \zeta_2^2
\nn\\
&
        -\frac{512}{3} x^2 \zeta_3
        -\frac{128}{3} (4+5 x) \zeta_3
\Big)
\bigg]
%
%
\nn\\
& 
+ C_F n_h^2 T_F^2    \bigg[ 
\iomx^6 \Big(
        -\frac{64}{9} \big(
                18
                +\ixo
                -75 x
                +176 x^2
                -75 x^3
                +18 x^4
                +x^5
        \big) H_{0,0,1}(x)
\nn\\
&
        +\frac{128}{9} \big(
                24-51 x+38 x^2+51 x^3-36 x^4+6 x^5\big) \zeta_3
\Big)
+\iomx^4 \Big(
        -\frac{32}{27} \big(
                294-397 x+72 x^2
\nn\\
&
                +127 x^3\big)
        -\frac{16}{81} \big(
                2632
                +317 \ixo
                -4746 x
                +2632 x^2
                +317 x^3
        \big) H_1(x)
        +512 x H_{0,0,0,1}(x)
\nn\\
&
        +1024 x H_{1,0,0,1}(x)
        +\frac{1536}{5} x \zeta_2^2
        -1024 x H_1(x) \zeta_3
\Big)
-\frac{32 \fxmo \iomx^5}{27} \big(
        25+318 x-494 x^2
\nn\\
&
        +318 x^3+25 x^4\big) H_{0,1}(x)
+\frac{1024}{45} \iomx \zeta_2
\bigg] \,.
\end{align}

\begin{align}
 {\mathcal{G}_3^{(3)}} &= C_F n_l^2 T_F^2  \bigg[
\ixtw \Big(
        -\frac{2848}{81} H_1(x)
        -\frac{832}{27} H_{0,1}(x)
        -\frac{1664}{27} H_{1,1}(x)
\Big)
+\ixo \Big(
        \frac{6496}{81}
\nn\\
&
        +\fxtnt \Big(
                -\frac{64}{9} H_{0,0,1}(x)
                -\frac{128}{9} H_{0,1,1}(x)
                -\frac{128}{9} H_{1,0,1}(x)
                -\frac{256}{9} H_{1,1,1}(x)
                -\frac{128}{9} H_1(x) \zeta_2
        \Big)
\nn\\
&
        +\frac{880}{9} H_1(x)
        +\frac{544}{9} H_{0,1}(x)
        +\frac{1088}{9} H_{1,1}(x)
        +\frac{256}{9} \zeta_2
\Big)
\bigg]
%
\nn\\
&
+ C_F^2 n_l T_F  \bigg[ 
\iomx \Big(
        \frac{1}{81} (-10049+10265 \ixo)
        +\frac{1}{162} \big(
                -18363+13549 \ixo-1522 \ixtw\big) H_1(x)
\Big)
\nn\\
&
+\iomx^2 \Big(
        -\frac{8}{27} \big(
                2788
                +7663 \ixo
                -2096 \ixtw
                +384 \ixthr
                +135 x
        \big) H_{1,1}(x)
        -\frac{64}{3} \big(
                62
                +41 \ixo
\nn\\
&
                -14 \ixtw
                +5 \ixthr
                -13 x
        \big) H_{1,1,1}(x)
\Big)
+\fxtnt \Big(
        -\frac{640}{3} \ixo H_{1,1,1,1}(x)
        -\frac{64}{3} \ixo H_{1,2,1,1}(x)
\nn\\
&
        -64 \ixo H_{1,2}(x) \zeta_2
\Big)
+\fxmo \Big(
        \iomx^3 \Big(
                \big(
                        -2+11 x-6 x^2+3 x^3
                \big)
\Big(-\frac{128}{3} \ixo H_{1,-1,0,1}(x)
\nn\\
&
                        -\frac{160}{3} \ixo H_{-1,0,0,1}(x)
                        -64 \ixo H_{-1,0,1,1}(x)
                        -\frac{128}{3} \ixo H_{-1,1,0,1}(x)
                        +\frac{128}{3} \ixo H_{-1,-1,0,1}(x)
\nn\\
&
                        +\Big(
                                64 \ixo \ln (2) H_{-1}(x)
                                -\frac{64}{3} \ixo H_{1,-1}(x)
                                +\frac{128}{3} \ixo H_{-1,1}(x)
                                +\frac{64}{3} \ixo H_{-1,-1}(x)
                        \Big) \zeta_2
\nn\\
&
                        -16 \ixo H_{-1}(x) \zeta_3
                \Big)
                +\frac{32}{9} \big(
                        -385
                        +70 \ixo
                        +234 x
                        -129 x^2
                \big) H_{-1,0,1}(x)
                +\frac{16}{9} \big(
                        -385
                        +70 \ixo
\nn\\
&
                        +234 x
                        -129 x^2
                \big) H_{-1}(x) \zeta_2
        \Big)
        +640 \iomx^4 x H_{0,0,1}(x) \zeta_2
\Big)
+\iomx^3 \Big(
        -
        \frac{8}{9} \big(
                -2499
                +343 \ixo
                +114 \ixtw
\nn\\
&
                -586 x
                -330 x^2
        \big) H_{0,1}(x)
        +\frac{8}{9} \big(
                2983
                -173 \ixo
                -18 \ixtw
                +12 \ixthr
                +208 x
                +72 x^2
        \big) H_{1,0,1}(x)
\nn\\
&
        +\frac{16}{3} \big(
                -32
                -96 \ixo
                +16 \ixtw
                -87 x
                +27 x^2
        \big) H_{1,0,0,1}(x)
        -\frac{32}{3} \big(
                207
                -29 \ixo
                +10 \ixtw
                -71 x
\nn\\
&
                +27 x^2
        \big) H_{1,0,1,1}(x)
        -\frac{32}{3} \big(
                -111
                -34 \ixo
                +4 \ixtw
                -27 x
                +6 x^2
        \big) H_{1,1,0,1}(x)
        +\big(
                -\frac{8}{9} \big(
                        -4824
\nn\\
&
                        +418 \ixo
                        +24 \ixtw
                        -3059 x
                        +535 x^2
                \big)
                +128 \ln (2) \big(
                        10-7 x+3 x^2\big) H_1(x)
                +\frac{4}{9} \big(
                        13885
                        +69 \ixo
\nn\\
&
                        +34 \ixtw
                        +48 \ixthr
                        +385 x
                        +147 x^2
                \big) H_1(x)
                +8 \big(
                        527
                        +19 \ixo
                        +10 \ixtw
                        -19 x
                        +15 x^2
                \big) H_{1,1}(x)
        \big) \zeta_2
\nn\\
&
        -\frac{4}{9} \big(
                615
                -753 \ixo
                +146 \ixtw
                -815 x
                +327 x^2
        \big) H_1(x) \zeta_3
\Big)
+\iomx^4 \Big(
        \frac{64}{3} a_4 \big(
                -46+82 x
\nn\\
&
                -64 x^2+19 x^3\big)
        +\big(
                -30+34 x-16 x^2+3 x^3
        \big)
\Big(\frac{64}{3} H_{2,1,1,1}(x)
                -\frac{32}{3} H_{2,2,1,1}(x)
\nn\\
&
                +\big(
                        32 \ln (2) H_2(x)
                        +32 H_{2,1}(x)
                        -32 H_{2,2}(x)
                \big) \zeta_2
                +
                \frac{56}{3} H_2(x) \zeta_3
        \Big)
\nn\\
&
        +(1+2 x) \Big(
                -\frac{128}{3} H_{0,0,1,0,1}(x)
                +\frac{128}{3} H_{0,1}(x) \zeta_3
        \Big)
        +(2+x) \Big(
                -\frac{512}{3} H_{0,0,-1,0,1}(x)
\nn\\
&
                -\frac{512}{3} H_{0,1,-1,0,1}(x)
                -\frac{640}{3} H_{0,-1,0,0,1}(x)
                -256 H_{0,-1,0,1,1}(x)
                -\frac{512}{3} H_{0,-1,1,0,1}(x)
\nn\\
&
                +\frac{512}{3} H_{0,-1,-1,0,1}(x)
                +\Big(
                         \ln (2) \big(
                                256 H_{0,1}(x)
                                +256 H_{0,-1}(x)
                        \big)
                        -\frac{256}{3} H_{0,0,-1}(x)
\nn\\
&
                        -\frac{256}{3} H_{0,1,-1}(x)
                        +\frac{512}{3} H_{0,-1,1}(x)
                        +\frac{256}{3} H_{0,-1,-1}(x)
                \Big) \zeta_2
                -64 H_{0,-1}(x) \zeta_3
        \Big)
\nn\\
&
        +\frac{16}{9} \big(
                320
                -295 \ixo
                +2 \ixtw
                -526 x
                +643 x^2
                -207 x^3
        \big) H_{0,0,1}(x)
        -\frac{16}{9} \big(
                1356
                +549 \ixo
\nn\\
&
                -86 \ixtw
                +24 \ixthr
                -2316 x
                +521 x^2
                -111 x^3
        \big) H_{0,1,1}(x)
        +\frac{16}{9} \big(
                -276
                +24 \ixo
                +334 x
\nn\\
&
                -181 x^2
                +36 x^3
        \big) H_{2,1,1}(x)
        +\frac{32}{9} \big(
                49
                -114 \ixo
                +12 \ixtw
                -88 x
                +213 x^2
                -54 x^3
        \big) H_{0,0,0,1}(x)
\nn\\
&
        -\frac{32}{3} \big(
                304
                +10 \ixo
                +4 \ixtw
                -80 x
                -18 x^2
                -3 x^3
        \big) H_{0,0,1,1}(x)
        +
        \frac{16}{3} \big(
                220
                -74 \ixo
                +12 \ixtw
                -5 x
\nn\\
&
                +46 x^2
                -9 x^3
        \big) H_{0,1,0,1}(x)
        -\frac{32}{3} \big(
                436
                -45 \ixo
                +14 \ixtw
                -542 x
                +170 x^2
                -51 x^3
        \big) H_{0,1,1,1}(x)
\nn\\
&
        +\frac{128}{9} \big(
                -13
                -33 \ixo
                +6 \ixtw
                -29 x
                -9 x^2
                +9 x^3
        \big) H_{0,-1,0,1}(x)
        -\frac{128}{3} (14+13 x) H_{0,0,0,0,1}(x)
\nn\\
&
        -\frac{128}{3} (34+29 x) H_{0,0,0,1,1}(x)
        -\frac{256}{3} (26+19 x) H_{0,0,1,1,1}(x)
        -\frac{64}{3} (20+19 x) H_{0,1,0,0,1}(x)
\nn\\
&
        -\frac{512}{3} (5+4 x) H_{0,1,0,1,1}(x)
        +\frac{64}{3} (46+35 x) H_{0,1,1,0,1}(x)
        +\Big(
                \frac{64}{3} \ln^2 (2) \big(
                        22-10 x-8 x^2+5 x^3\big)
\nn\\
&
                +\frac{16 \ln (2) }{3} \big(
                        476
                        +24 \ixo
                        -910 x
                        +685 x^2
                        -212 x^3
                \big)
                +\frac{16}{3} \big(
                        -276
                        +24 \ixo
                        +334 x
                        -181 x^2
\nn\\
&
                        +36 x^3
                \big) H_2(x)
                +\frac{4}{9} \big(
                        7768
                        -1731 \ixo
                        +402 \ixtw
                        -478 x
                        -1050 x^2
                        +369 x^3
                \big) H_{0,1}(x)
\nn\\
&
                +\frac{64}{9} \big(
                        -13
                        -33 \ixo
                        +6 \ixtw
                        -29 x
                        -9 x^2
                        +9 x^3
                \big) H_{0,-1}(x)
                +64 (38+31 x) H_{0,1,1}(x)
\nn\\
&
                -\frac{32}{3} (346+275 x) \zeta_3
        \Big) \zeta_2
        -\frac{8}{45} \big(
                1096+15884 x-8430 x^2+2247 x^3\big) \zeta_2^2
\nn\\
&
        +\frac{4}{9} \big(
                -1604
                +74 \ixo
                +4358 x
                -3841 x^2
                +1202 x^3
        \big) \zeta_3
        +16 (62+47 x) \zeta_5
\Big)
\bigg]
%
%
\nn\\
&
+ C_F C_A n_l T_F  \bigg[ 
\fxmo \iomx^3 \Big(
        \big(
                -2+11 x-6 x^2+3 x^3
        \big)
\Big(
                \frac{64}{3} \ixo H_{1,-1,0,1}(x)
                +\frac{80}{3} \ixo H_{-1,0,0,1}(x)
\nn\\
&
                +32 \ixo H_{-1,0,1,1}(x)
                +\frac{64}{3} \ixo H_{-1,1,0,1}(x)
                -\frac{64}{3} \ixo H_{-1,-1,0,1}(x)
                +\Big(
                        -32 \ixo \ln (2) H_{-1}(x)
\nn\\
&
                        +\frac{32}{3} \ixo H_{1,-1}(x)
                        -\frac{64}{3} \ixo H_{-1,1}(x)
                        -\frac{32}{3} \ixo H_{-1,-1}(x)
                \Big) \zeta_2
                +8 \ixo H_{-1}(x) \zeta_3
        \Big)
        -\frac{16}{9} \big(
                -385
\nn\\
&
                +70 \ixo
                +234 x
                -129 x^2
        \big) H_{-1,0,1}(x)
        -\frac{8}{9} \big(
                -385
                +70 \ixo
                +234 x
                -129 x^2
        \big) H_{-1}(x) \zeta_2
\Big)
\nn\\
&
+\iomx \Big(
        -\frac{8}{81} (-6209+7046 \ixo)
        +\frac{16}{81} \big(
                4095-7210 \ixo+1954 \ixtw\big) H_1(x)
\Big)
+\iomx^2 \Big(
        \frac{4}{27} \big(
                8288
\nn\\
&
                -12457 \ixo
                +2534 \ixtw
                -4557 x
        \big) H_{1,1}(x)
        +\frac{8}{9} \big(
                676
                -1181 \ixo
                +262 \ixtw
                -405 x
        \big) H_{1,1,1}(x)
\Big)
\nn\\
&
+\fxtnt \Big(
        \frac{32}{3} \ixo H_{1,2,1,1}(x)
        +32 \ixo H_{1,2}(x) \zeta_2
\Big)
+\iomx^3 \Big(
        \frac{4}{27} \big(
                14049
                -9072 \ixo
                +1564 \ixtw
                -2140 x
\nn\\
&
                +1791 x^2
        \big) H_{0,1}(x)
        +
        \frac{4}{9} \big(
                3393
                -1495 \ixo
                +334 \ixtw
                -813 x
                +519 x^2
        \big) H_{1,0,1}(x)
\nn\\
&
        -\frac{16}{3} \big(
                101
                -12 \ixo
                +4 \ixtw
                -2 x
                +12 x^2
        \big) H_{1,0,0,1}(x)
        -\frac{32}{3} \big(
                32
                +4 \ixo
                +21 x
                -3 x^2
        \big) H_{1,0,1,1}(x)
\nn\\
&
        +\frac{8}{3} \big(
                134
                +16 \ixo
                +33 x
                +15 x^2
        \big) H_{1,1,0,1}(x)
        +\Big(
                \frac{4}{9} \big(
                        5724
                        +212 \ixo
                        +879 x
                        +913 x^2
                \big)
\nn\\
&
                -64 \ln (2) \big(
                        10-7 x+3 x^2\big) H_1(x)
                +\frac{8}{9} \big(
                        2924
                        +102 \ixo
                        +36 \ixtw
                        +364 x
                        +249 x^2
                \big) H_1(x)
\nn\\
&
                +\frac{8}{3} \big(
                        234
                        +192 \ixo
                        -32 \ixtw
                        +287 x
                        -15 x^2
                \big) H_{1,1}(x)
        \Big) \zeta_2
        +\frac{16}{3} \big(
                98
                -25 \ixo
                +6 \ixtw
                -36 x
\nn\\
&
                +18 x^2
        \big) H_1(x) \zeta_3
\Big)
+\iomx^4 \Big(
        -\frac{32}{3} a_4 \big(
                -46+82 x-64 x^2+19 x^3\big)
        +\big(
                -30+34 x-16 x^2
\nn\\
&
                +3 x^3
        \big)
\Big(-\frac{32}{3} H_{2,1,1,1}(x)
                +\frac{16}{3} H_{2,2,1,1}(x)
                +\big(
                        -16 \ln (2) H_2(x)
                        -16 H_{2,1}(x)
                        +16 H_{2,2}(x)
                \big) \zeta_2
\nn\\
&
                -\frac{28}{3} H_2(x) \zeta_3
        \Big)
        +(2+x) \Big(
                \frac{256}{3} H_{0,0,-1,0,1}(x)
                +
                \frac{256}{3} H_{0,1,-1,0,1}(x)
                +\frac{320}{3} H_{0,-1,0,0,1}(x)
\nn\\
&
                +128 H_{0,-1,0,1,1}(x)
                +\frac{256}{3} H_{0,-1,1,0,1}(x)
                -\frac{256}{3} H_{0,-1,-1,0,1}(x)
                +\Big(
                        \ln (2) \big(
                                -128 H_{0,1}(x)
\nn\\
&
                                -128 H_{0,-1}(x)
                        \big)
                        +\frac{128}{3} H_{0,0,-1}(x)
                        +\frac{128}{3} H_{0,1,-1}(x)
                        -\frac{256}{3} H_{0,-1,1}(x)
                        -\frac{128}{3} H_{0,-1,-1}(x)
                \Big) \zeta_2
\nn\\
&
                +32 H_{0,-1}(x) \zeta_3
        \Big)
        +\frac{4}{9} \big(
                -848
                -1240 \ixo
                +176 \ixtw
                +1831 x
                +42 x^2
                +165 x^3
        \big) H_{0,0,1}(x)
\nn\\
&
        +\frac{8}{9} \big(
                860
                -1066 \ixo
                +140 \ixtw
                -779 x
                +1007 x^2
                -225 x^3
        \big) H_{0,1,1}(x)
        -\frac{8}{9} \big(
                -276
                +24 \ixo
\nn\\
&
                +334 x
                -181 x^2
                +36 x^3
        \big) H_{2,1,1}(x)
        -\frac{8}{9} \big(
                1293
                +48 \ixo
                -395 x
                +143 x^2
                -108 x^3
        \big) H_{0,0,0,1}(x)
\nn\\
&
        -\frac{16}{9} \big(
                715
                +48 \ixo
                -447 x
                +23 x^2
                -45 x^3
        \big) H_{0,0,1,1}(x)
        -\frac{8}{9} \big(
                287
                +12 \ixo
                -696 x
                +67 x^2
\nn\\
&
                -45 x^3
        \big) H_{0,1,0,1}(x)
        -\frac{32}{3} \big(
                65
                +8 \ixo
                -32 x
                -35 x^2
                +3 x^3
        \big) H_{0,1,1,1}(x)
        -\frac{64}{9} \big(
                -13
                -33 \ixo
\nn\\
&
                +6 \ixtw
                -29 x
                -9 x^2
                +9 x^3
        \big) H_{0,-1,0,1}(x)
        -
        \frac{16}{3} \big(
                78+59 x+16 x^2\big) H_{0,0,0,0,1}(x)
        -\frac{32}{3} \big(
                66+53 x
\nn\\
&
                +16 x^2\big) H_{0,0,0,1,1}(x)
        -\frac{16}{3} \big(
                40+25 x+4 x^2\big) H_{0,0,1,0,1}(x)
        -\frac{64}{3} \big(
                18+19 x+8 x^2\big) H_{0,0,1,1,1}(x)
\nn\\
&
        -\frac{16}{3} \big(
                56+43 x+12 x^2\big) H_{0,1,0,0,1}(x)
        -\frac{64}{3} \big(
                12+11 x+4 x^2\big) H_{0,1,0,1,1}(x)
        +\frac{16}{3} \big(
                42+41 x
\nn\\
&
                +16 x^2\big) H_{0,1,1,0,1}(x)
        +\Big(
                -\frac{32}{3}  \ln^2 (2) \big(
                        22-10 x-8 x^2+5 x^3\big)
                -\frac{8 \ln (2)}{3} \big(
                        476
                        +24 \ixo
                        -910 x
\nn\\
&
                        +685 x^2
                        -212 x^3
                \big)
                +\frac{8}{3} \big(
                         276
                        -24 \ixo
                        -334 x
                        +181 x^2
                        -36 x^3
                \big) H_2(x)
                -\frac{8}{9} \big(
                        -773
                        -324 \ixo
\nn\\
&
                        +48 \ixtw
                        -484 x
                        -81 x^2
                        +99 x^3
                \big) H_{0,1}(x)
                -\frac{32}{9} \big(
                        -13
                        -33 \ixo
                        +6 \ixtw
                        -29 x
                        -9 x^2
\nn\\
&
                        +9 x^3
                \big) H_{0,-1}(x)
                +\frac{16}{3} \big(
                        76+63 x+20 x^2\big) H_{0,0,1}(x)
                +16 \big(
                        50+45 x+16 x^2\big) H_{0,1,1}(x)
\nn\\
&
                -\frac{16}{3} \big(
                        168+169 x+68 x^2\big) \zeta_3
        \Big) \zeta_2
        +\frac{4}{45} \big(
                -8345-2368 x-5217 x^2+1212 x^3\big) \zeta_2^2
\nn\\
&
        +\Big(
                -\frac{2}{9} \big(
                         72 \ixo
                        -3132
                        +5582 x
                        -3641 x^2
                        +1308 x^3
                \big)
                +\frac{16}{3} \big(
                        40+25 x+4 x^2\big) H_{0,1}(x)
        \Big) \zeta_3
\nn\\
&
        +\frac{32}{3} \big(
                10+15 x+8 x^2\big) \zeta_5
\Big)
\bigg]
%
\nn\\
&
+ C_F n_l n_h T_F^2  \bigg[ 
\fxmo \iomx^3 \Big(
        -\frac{128}{9} \big(
                -11
                +2 \ixo
                -10 x
                -3 x^2
        \big) H_{0,1,1}(x)
        -\frac{128}{9} \big(
                -11
                +2 \ixo
\nn\\
&
                -10 x
                -3 x^2
        \big) H_{1,0,1}(x)
        +\frac{128}{9} \big(
                -11
                +2 \ixo
                -10 x
                -3 x^2
        \big) H_1(x) \zeta_2
\Big)
+\iomx^2 \Big(
        \frac{32}{81} (-1256
\nn\\
&
        +406 \ixo
        +109 x
        )
        -\frac{32}{81} \big(
                862
                -1847 \ixo
                +178 \ixtw
                -495 x
        \big) H_1(x)
        -\frac{64}{27} \big(
                -124
                -223 \ixo
\nn\\
&
                +26 \ixtw
                -51 x
        \big) H_{1,1}(x)
\Big)
+\iomx^3 \Big(
        -\frac{64}{27} \big(
                39
                -189 \ixo
                +26 \ixtw
                +78 x
        \big) H_{0,1}(x)
        -\frac{128}{27} \big(
                114
\nn\\
&
                +6 \ixo
                -25 x
                +45 x^2
        \big) \zeta_2
\Big)
+\iomx^4 \Big(
        (2+x) \Big(
                -\frac{512}{3} H_{0,0,0,1}(x)
                -\frac{512}{3} H_{0,0,1,1}(x)
\nn\\
&
                -\frac{512}{3} H_{0,1,0,1}(x)
                +\frac{512}{3} H_{0,1}(x) \zeta_2
                -\frac{512}{15} \zeta_2^2
        \Big)
        -\frac{128}{9} \big(
                26
                -11 \ixo
                +2 \ixtw
                -3 x
\nn\\
&
                +22 x^2
                +3 x^3
        \big) H_{0,0,1}(x)
        +\frac{128}{9} \big(
                38-11 x+12 x^2\big) \zeta_3
\Big)
\bigg]
%
\nn\\
&
+ C_F n_h^2 T_F^2  \bigg[ 
\iomx^4 \Big(
        \frac{32}{81} \big(
                940
                +203 \ixo
                -2103 x
                +1342 x^2
                -94 x^3
        \big)
        +(2+x) \Big(
                -\frac{512}{3} H_{0,0,0,1}(x)
\nn\\
&
                -\frac{1024}{3} H_{1,0,0,1}(x)
                -\frac{512}{5} \zeta_2^2
                +\frac{1024}{3} H_1(x) \zeta_3
        \Big)
        -\frac{16}{81} \big(
                3972
                -3199 \ixo
                +178 \ixtw
                -1858 x
\nn\\
&
                +250 x^2
                -495 x^3
        \big) H_1(x)
\Big)
+\iomx^6 \Big(
        -\frac{64}{9} \big(
                52
                -15 \ixo
                +2 \ixtw
                -143 x
                +42 x^2
                +5 x^3
                -4 x^4
\nn\\
&
                -3 x^5
        \big) H_{0,0,1}(x)
        +\frac{128}{9} \big(
                -46+133 x-228 x^2+151 x^3-48 x^4+6 x^5\big) \zeta_3
\Big)
\nn\\
&
-\frac{32 \fxmo \iomx^5}{27} \big(
        -327
        +26 \ixo
        +266 x
        -90 x^2
        -16 x^3
        -51 x^4
\big) H_{0,1}(x)
-\frac{1024}{45} \iomx \zeta_2
\bigg]   \,.
\end{align}

\begin{align}
 {\mathcal{S}^{(3)}} &= C_F n_l^2 T_F^2  \bigg[
-\frac{59891}{6561}
+(\ixo (3+5 x)) \Big(
        -\frac{64}{27} H_{0,0,1}(x)
        -\frac{128}{27} H_{0,1,1}(x)
        -\frac{128}{27} H_{1,0,1}(x)
\nn\\
&
        -\frac{256}{27} H_{1,1,1}(x)
        -\frac{128}{27} H_1(x) \zeta_2
\Big)
+\ixo \Big(
        -\frac{2336}{81} H_1(x)
        -\frac{512}{27} H_{0,1}(x)
        -\frac{1024}{27} H_{1,1}(x)
\Big)
\nn\\
&
-\frac{3712}{729} H_1(x)
-\frac{800}{81} H_{0,1}(x)
-\frac{1600}{81} H_{1,1}(x)
-\frac{64}{9} H_{0,0,0,1}(x)
-\frac{128}{9} H_{0,0,1,1}(x)
\nn\\
&
-\frac{128}{9} H_{0,1,0,1}(x)
-\frac{256}{9} H_{0,1,1,1}(x)
-\frac{128}{9} H_{1,0,0,1}(x)
-\frac{256}{9} H_{1,0,1,1}(x)
-\frac{256}{9} H_{1,1,0,1}(x)
\nn\\
&
-\frac{512}{9} H_{1,1,1,1}(x)
+\Big(
        -\frac{4232}{81}
        -\frac{128}{9} H_{0,1}(x)
        -\frac{256}{9} H_{1,1}(x)
\Big) \zeta_2
-\frac{1768}{135} \zeta_2^2
\nn\\
&
-\frac{64}{243} \zeta_3 \big(
        130+63 H_1(x)\big)
\bigg]
%
%
\nn\\
&
+ C_F^2 n_l T_F  \bigg[
\frac{159271}{972}
+\fxmosq \iomx \Big(
        \frac{128}{3} x H_{1,-1,0,1}(x)
        +\frac{160}{3} x H_{-1,0,0,1}(x)
        +64 x H_{-1,0,1,1}(x)
\nn\\
&
        +\frac{128}{3} x H_{-1,1,0,1}(x)
        -\frac{128}{3} x H_{-1,-1,0,1}(x)
        +\Big(
                -64 \ln (2) x H_{-1}(x)
                +\frac{64}{3} x H_{1,-1}(x)
\nn\\
&
                -\frac{128}{3} x H_{-1,1}(x)
                -\frac{64}{3} x H_{-1,-1}(x)
        \Big) \zeta_2
        +16 x H_{-1}(x) \zeta_3
\Big)
-\ixtw \Big(
         \frac{512}{9} H_{1,1}(x)
        +\frac{160}{3} H_{1,1,1}(x)
\Big)
\nn\\
&
+\ixo \big(
        -\frac{2561}{162} H_1(x)
        +\frac{5312}{27} H_{1,1}(x)
        +32 H_{1,1,1}(x)
        +\frac{128}{3} H_{1,0,0,1}(x)
\big)
\nn\\
&
+\fxmo \Big(
        \iomx \Big(
                \frac{32}{9} (41+20 x) H_{-1,0,1}(x)
                +\frac{16}{9} (41+20 x) H_{-1}(x) \zeta_2
        \Big)
        -\frac{640}{3} H_{1,1,1,1}(x)
\Big)
\nn\\
&
+\fxtnt \Big(
        -\frac{32}{3} H_{1,2,1,1}(x)
        -32 H_{1,2}(x) \zeta_2
\Big)
+\iomx \Big(
        \ixtw \Big(
                \frac{16}{3} H_{1,0,1}(x)
                +\frac{32}{3} H_1(x) \zeta_2
        \Big)
\nn\\
&
        +\ixo \Big(
                -\frac{352}{9} H_{0,1}(x)
                +16 H_{1,0,1}(x)
                -\frac{160}{3} H_{1,0,1,1}(x)
                -\frac{64}{3} H_{1,1,0,1}(x)
                +\Big(
                        -\frac{32}{3}
                        +
                        \frac{614}{9} H_1(x)
\nn\\
&
                        +40 H_{1,1}(x)
                \Big) \zeta_2
                -\frac{292}{9} H_1(x) \zeta_3
        \Big)
        -\frac{1}{162} (33377+13423 x) H_{0,1}(x)
        -\frac{8}{27} (649-19 x) H_{1,0,1}(x)
\nn\\
&
        +\frac{32}{9} (31+26 x) H_{1,0,1,1}(x)
        +\frac{16}{9} (-71+47 x) H_{1,1,0,1}(x)
        +\Big(
                \frac{1}{648} (213559-139255 x)
\nn\\
&
                +32 \ln (2) (-9+x) H_1(x)
                +\frac{2}{27} (-5701+2548 x) H_1(x)
                +\frac{4}{9} (-941+419 x) H_{1,1}(x)
        \Big) \zeta_2
\nn\\
&
        +\frac{4}{9} (119+74 x) H_1(x) \zeta_3
\Big)
+\iomx^2 \Big(
        \frac{32}{3} a_4 \big(
                52-60 x+11 x^2\big)
        +\ixo \Big(
                \frac{112}{9} H_{0,0,1}(x)
\nn\\
&
                +\frac{304}{9} H_{0,1,1}(x)
                +\frac{64}{3} H_{0,0,0,1}(x)
                -\frac{64}{3} H_{0,0,1,1}(x)
                +32 H_{0,1,0,1}(x)
                -\frac{224}{3} H_{0,1,1,1}(x)
\nn\\
&
                +\frac{128}{3} H_{0,-1,0,1}(x)
                +\Big(
                        \frac{268}{3} H_{0,1}(x)
                        +\frac{64}{3} H_{0,-1}(x)
                \Big) \zeta_2
        \Big)
        +\frac{8}{9} \big(
                -123+120 x+16 x^2\big) H_{0,0,1}(x)
\nn\\
&
        -\frac{64}{3} \ixtw H_{0,1,1}(x)
        +\frac{8}{27} \big(
                526-986 x+337 x^2\big) H_{0,1,1}(x)
        +\frac{8}{3} (-2+x) (-26+17 x) H_{2,1,1}(x)
\nn\\
&
        +\frac{16}{9} \big(
                -6-44 x+47 x^2\big) H_{0,0,0,1}(x)
        -
        \frac{16}{9} \big(
                -71+14 x+6 x^2\big) H_{0,0,1,1}(x)
\nn\\
&
        -\frac{128}{9} \big(
                4-3 x+2 x^2\big) H_{0,1,0,1}(x)
        -\frac{32}{3} \big(
                -28+11 x+13 x^2\big) H_{0,1,1,1}(x)
\nn\\
&
        +\frac{32}{9} \big(
                11-34 x+5 x^2\big) H_{0,-1,0,1}(x)
        +32 (-2+x)^2 H_{2,1,1,1}(x)
        -16 (-2+x)^2 H_{2,2,1,1}(x)
\nn\\
&
        +\frac{64}{3} \big(
                3-2 x+2 x^2\big) H_{0,0,0,0,1}(x)
        +\frac{128}{3} x H_{0,0,0,1,1}(x)
        +\frac{64}{3} \big(
                3+x^2\big) H_{0,0,0,1,1}(x)
\nn\\
&
        +\frac{16}{3} \big(
                8-12 x+7 x^2\big) H_{0,0,1,0,1}(x)
        +\frac{32}{3} \big(
                1+14 x-3 x^2\big) H_{0,0,1,1,1}(x)
\nn\\
&
        +\frac{16}{3} \big(
                14-16 x+11 x^2\big) H_{0,1,0,0,1}(x)
        -\frac{32}{3} \big(
                1-10 x+3 x^2\big) H_{0,1,0,1,1}(x)
\nn\\
&
        -\frac{64}{3} \big(
                3-2 x+2 x^2\big) H_{0,1,1,0,1}(x)
        +\Big(
                \frac{32}{3}  \ln^2 (2) \big(
                        8-12 x+x^2\big)
                +\ln (2) \Big(
                        -\frac{8}{3} \big(
                                560-732 x
\nn\\
&
                                +199 x^2\big)
                        +48 (-2+x)^2 H_2(x)
                \Big)
                +8 (-2+x) (-26+17 x) H_2(x)
\nn\\
&
                -\frac{22}{9} \big(
                        73-32 x+37 x^2\big) H_{0,1}(x)
                +\frac{16}{9} \big(
                        11-34 x+5 x^2\big) H_{0,-1}(x)
                +48 (-2+x)^2 H_{2,1}(x)
\nn\\
&
                -48 (-2+x)^2 H_{2,2}(x)
                +\frac{4}{3} \big(
                        27-134 x+47 x^2\big) H_{0,0,1}(x)
                -8 \big(
                        11+10 x+3 x^2\big) H_{0,1,1}(x)
\nn\\
&
                +
                \frac{1}{9} \big(
                        2539-1814 x+1723 x^2\big) \zeta_3
        \Big) \zeta_2
        +\frac{1}{270} \big(
                -81187+107942 x-9547 x^2\big) \zeta_2^2
\nn\\
&
        +\Big(
                \frac{2}{81} \big(
                        16924-23660 x+6007 x^2\big)
                +28 (-2+x)^2 H_2(x)
                -\frac{4}{9} \big(
                        97-146 x+85 x^2\big) H_{0,1}(x)
        \Big) \zeta_3
\nn\\
&
        +\frac{64}{9} \big(
                1-14 x+4 x^2\big) \zeta_5
\Big)
+\frac{397}{9} H_1(x)
+\frac{6295}{81} H_{1,1}(x)
+\frac{704}{9} H_{1,1,1}(x)
+\frac{688}{9} H_{1,0,0,1}(x)
\nn\\
&
+\frac{128}{3} H_{0,0,-1,0,1}(x)
-\frac{640}{3} H_{0,1,1,1,1}(x)
-\frac{64}{3} H_{0,1,2,1,1}(x)
+\frac{128}{3} H_{0,1,-1,0,1}(x)
\nn\\
&
+\frac{160}{3} H_{0,-1,0,0,1}(x)
+64 H_{0,-1,0,1,1}(x)
+\frac{128}{3} H_{0,-1,1,0,1}(x)
-\frac{128}{3} H_{0,-1,-1,0,1}(x)
\nn\\
&
+128 H_{1,0,0,0,1}(x)
+\frac{128}{3} H_{1,0,1,0,1}(x)
-192 H_{1,0,1,1,1}(x)
+\frac{352}{3} H_{1,1,0,0,1}(x)
-128 H_{1,1,0,1,1}(x)
\nn\\
&
-\frac{256}{3} H_{1,1,1,0,1}(x)
-\frac{1280}{3} H_{1,1,1,1,1}(x)
-\frac{64}{3} H_{1,1,2,1,1}(x)
+\Big(
        \ln (2) \big(
                 64 H_{1,1}(x)
                -64 H_{0,-1}(x)
        \big)
\nn\\
&
        +\frac{64}{3} H_{0,0,-1}(x)
        -64 H_{0,1,2}(x)
        +\frac{64}{3} H_{0,1,-1}(x)
        -\frac{128}{3} H_{0,-1,1}(x)
        -\frac{64}{3} H_{0,-1,-1}(x)
        -\frac{8}{3} H_{1,0,1}(x)
\nn\\
&
        -\frac{208}{3} H_{1,1,1}(x)
        -64 H_{1,1,2}(x)
\Big) \zeta_2
+\frac{488}{15} H_1(x) \zeta_2^2
+\Big(
        16 H_{0,-1}(x)
        -\frac{728}{9} H_{1,1}(x)
\Big) \zeta_3
\bigg]
%
%
\nn\\
&
+ C_F C_A n_l T_F  \bigg[
\frac{854630}{6561}
+(\fxmo \iomx (41+20 x)) \Big(
        -\frac{16}{9} H_{-1,0,1}(x)
        -\frac{8}{9} H_{-1}(x) \zeta_2
\Big)
\nn\\
&
+\fxmosq \iomx \Big(
        -\frac{64}{3} x H_{1,-1,0,1}(x)
        -\frac{80}{3} x H_{-1,0,0,1}(x)
        -32 x H_{-1,0,1,1}(x)
        -\frac{64}{3} x H_{-1,1,0,1}(x)
\nn\\
&
        +\frac{64}{3} x H_{-1,-1,0,1}(x)
        +\Big(
                32 \ln (2) x H_{-1}(x)
                -\frac{32}{3} x H_{1,-1}(x)
                +\frac{64}{3} x H_{-1,1}(x)
                +\frac{32}{3} x H_{-1,-1}(x)
        \Big) \zeta_2
\nn\\
&
        -8 x H_{-1}(x) \zeta_3
\Big)
+\ixo \Big(
        \frac{21908}{81} H_1(x)
        +\frac{5908}{27} H_{1,1}(x)
        +\frac{1048}{9} H_{1,1,1}(x)
\Big)
+\fxtnt \Big(
        \frac{16}{3} H_{1,2,1,1}(x)
\nn\\
&
        +16 H_{1,2}(x) \zeta_2
\Big)
+\iomx \Big(
        \ixo \Big(
                \frac{3656}{27} H_{0,1}(x)
                +\frac{668}{9} H_{1,0,1}(x)
                -\frac{32}{3} H_{1,0,0,1}(x)
                +16 H_1(x) \zeta_3
\nn\\
&
                +\Big(
                        \frac{16}{3} H_1(x)
                        -\frac{128}{3} H_{1,1}(x)
                \Big) \zeta_2
        \Big)
        +\frac{4}{81} (227-1844 x) H_{0,1}(x)
        -\frac{4}{27} (-649+1021 x) H_{1,0,1}(x)
\nn\\
&
        -\frac{32}{3} (-5+x) H_{1,0,0,1}(x)
        -\frac{32}{9} (-32+41 x) H_{1,0,1,1}(x)
        -\frac{4}{9} (-301+325 x) H_{1,1,0,1}(x)
\nn\\
&
        +\Big(
                \frac{1}{729} (7879-144283 x)
                -16 \ln (2) (-9+x) H_1(x)
                -\frac{4}{81} (-2137+4324 x) H_1(x)
\nn\\
&
                -\frac{4}{9} (-371+347 x) H_{1,1}(x)
        \Big) \zeta_2
        -\frac{4}{3} (29+11 x) H_1(x) \zeta_3
\Big)
+\iomx^2 \Big(
        -\frac{16}{3} a_4 \big(
                52-60 x+11 x^2\big)
\nn\\
&
        +\ixo \Big(
                \frac{352}{9} H_{0,0,1}(x)
                +\frac{560}{9} H_{0,1,1}(x)
                -\frac{64}{3} H_{0,-1,0,1}(x)
                +\Big(
                        -\frac{64}{3} H_{0,1}(x)
                        -\frac{32}{3} H_{0,-1}(x)
                \Big) \zeta_2
        \Big)
\nn\\
&
        +\frac{4}{27} \big(
                590-1399 x+464 x^2\big) H_{0,0,1}(x)
        +\frac{4}{27} \big(
                487-1892 x+1066 x^2\big) H_{0,1,1}(x)
\nn\\
&
        -\frac{4}{3} (-2+x) (-26+17 x) H_{2,1,1}(x)
        +\frac{4}{9} \big(
                174-152 x+x^2\big) H_{0,0,0,1}(x)
        +\frac{16}{9} \big(
                69-106 x
\nn\\
&
                +38 x^2\big) H_{0,0,1,1}(x)
        +\frac{4}{9} \big(
                183-358 x+146 x^2\big) H_{0,1,0,1}(x)
        +\frac{8}{9} \big(
                173-358 x+203 x^2\big) H_{0,1,1,1}(x)
\nn\\
&
        -\frac{16}{9} \big(
                11-34 x+5 x^2\big) H_{0,-1,0,1}(x)
        -16 (-2+x)^2 H_{2,1,1,1}(x)
        +8 (-2+x)^2 H_{2,2,1,1}(x)
\nn\\
&
        +\frac{32}{3} H_{0,0,0,0,1}(x)
        +\frac{64}{3} H_{0,0,0,1,1}(x)
        +\frac{8}{3} H_{0,0,1,0,1}(x)
        +
        \frac{64}{3} H_{0,0,1,1,1}(x)
        -\frac{32}{3} x^2 H_{0,1,0,0,1}(x)
\nn\\
&
        +\frac{32}{3} H_{0,1,0,1,1}(x)
        -\frac{32}{3} H_{0,1,1,0,1}(x)
        +\Big(
                -\frac{16}{3}  \ln^2 (2) \big(
                        8-12 x+x^2\big)
                +\ln (2) \Big(
                        \frac{4}{3} \big(
                                560-732 x
\nn\\
&
                                +199 x^2\big)
                        -24 (x-2)^2 H_2(x)
                \Big)
                -4 (x-2) (-26+17 x) H_2(x)
                +\frac{4}{3} \big(
                        18-48 x+43 x^2\big) H_{0,1}(x)
\nn\\
&
                -\frac{8}{9} \big(
                        11-34 x+5 x^2\big) H_{0,-1}(x)
                -24 (-2+x)^2 H_{2,1}(x)
                +24 (-2+x)^2 H_{2,2}(x)
\nn\\
&
                -\frac{8}{3} \big(
                        13-16 x+8 x^2\big) H_{0,0,1}(x)
                -\frac{32}{3} \big(
                        7-8 x+4 x^2\big) H_{0,1,1}(x)
                +\frac{4}{3} \big(
                        15+38 x-19 x^2\big) \zeta_3
        \Big) \zeta_2
\nn\\
&
        +\frac{2}{135} \big(
                6899-6886 x+1157 x^2\big) \zeta_2^2
        +\Big(
                \frac{1}{81} \big(
                        -10376+10348 x+757 x^2\big)
                -14 (x-2)^2 H_2(x)
\nn\\
&
                +\frac{8}{3} \big(
                        5-12 x+6 x^2\big) H_{0,1}(x)
        \Big) \zeta_3
        -\frac{4}{3} \big(
                59-102 x+51 x^2\big) \zeta_5
\Big)
+\frac{14998}{729} H_1(x)
+\frac{14284}{81} H_{1,1}(x)
\nn\\
&
+\frac{8456}{27} H_{1,1,1}(x)
+\frac{2816}{9} H_{1,1,1,1}(x)
-\frac{64}{3} H_{0,0,-1,0,1}(x)
-\frac{8}{3} \fxei
 \iomx^2 x H_{0,1,0,0,1}(x)
\nn\\
&
+\frac{32}{3} H_{0,1,2,1,1}(x)
-\frac{64}{3} H_{0,1,-1,0,1}(x)
-\frac{80}{3} H_{0,-1,0,0,1}(x)
-32 H_{0,-1,0,1,1}(x)
-\frac{64}{3} H_{0,-1,1,0,1}(x)
\nn\\
&
+\frac{64}{3} H_{0,-1,-1,0,1}(x)
-\frac{176}{3} H_{1,0,0,0,1}(x)
-\frac{128}{3} H_{1,0,0,1,1}(x)
-\frac{64}{3} H_{1,0,1,0,1}(x)
-64 H_{1,1,0,0,1}(x)
\nn\\
&
-\frac{32}{3} H_{1,1,0,1,1}(x)
+\frac{32}{3} H_{1,1,2,1,1}(x)
+\Big(
        \ln (2) \big(
                32 H_{0,-1}(x)
                -32 H_{1,1}(x)
        \big)
        -\frac{32}{3} H_{0,0,-1}(x)
\nn\\
&
        +32 H_{0,1,2}(x)
        -\frac{32}{3} H_{0,1,-1}(x)
        +\frac{64}{3} H_{0,-1,1}(x)
        +\frac{32}{3} H_{0,-1,-1}(x)
        -\frac{128}{3} H_{1,1,1}(x)
\nn\\
&
        +32 H_{1,1,2}(x)
\Big) \zeta_2
-\frac{152}{3} H_1(x) \zeta_2^2
+\big(
        -8 H_{0,-1}(x)
        +24 H_{1,1}(x)
\big) \zeta_3
\bigg]
%
%
\nn\\
&
+ C_F n_l n_h T_F^2  \bigg[
\fxmo \iomx^3 \big(
        3+11 x-11 x^2+5 x^3
\big)
\Big(-\frac{128}{27} H_{0,1,1}(x)
        -\frac{128}{27} H_{1,0,1}(x)
\nn\\
&
        +\frac{128}{27} H_1(x) \zeta_2
\Big)
+\iomx^3 \Big(
         \frac{64}{81} \big(
                11+153 x-99 x^2+31 x^3\big) H_{0,1}(x)
        -\frac{128}{27} \ixo (3+5 x) H_{0,0,1}(x)
\nn\\
&
        -\frac{1024}{27} \ixo H_{0,1}(x)
        -\frac{128}{27} x \big(
                9-6 x+5 x^2\big) H_{0,0,1}(x)
        -\frac{16}{81} \big(
                -317+1287 x-1047 x^2
\nn\\
&
                +269 x^3\big) \zeta_2
\Big)
+\iomx^2 \Big(
        -\frac{4}{729} \big(
                -7751-25970 x+12985 x^2\big)
        +\ixo \Big(
                -\frac{4672}{81} H_1(x)
\nn\\
&
                -\frac{1024}{27} H_{1,1}(x)
        \Big)
        +\frac{64}{81} \big(
                178-65 x+8 x^2\big) H_1(x)
        -\frac{512}{81} \big(
                13-14 x+7 x^2\big) H_{1,1}(x)
        -\frac{128}{3} \zeta_3
\Big)
\nn\\
&
-\frac{128}{9} H_{0,0,0,1}(x)
-\frac{128}{9} H_{0,0,1,1}(x)
-\frac{128}{9} H_{0,1,0,1}(x)
+\frac{128}{9} H_{0,1}(x) \zeta_2
-\frac{128}{45} \zeta_2^2
\bigg]
%
%
\nn\\
&
+ C_F n_h^2 T_F^2  \bigg[
\iomx^4 \Big(
        -\frac{2}{729} \big(
                69839-271148 x+388866 x^2-251708 x^3+62423 x^4\big)
\nn\\
&
        -\frac{2336}{81} \ixo H_1(x)
        -\frac{16}{243} \big(
                -865-332 x+1134 x^2-722 x^3+203 x^4\big) H_1(x)
\Big)
\nn\\
&
+\iomx^6 \Big(
        -\frac{64}{27} \ixo (3+5 x) H_{0,0,1}(x)
        -\frac{64 x}{27} \big(
                -21+9 x-39 x^2+57 x^3-27 x^4
\nn\\
&
                +5 x^5\big) H_{0,0,1}(x)
        +\frac{32}{27} \big(
                151-762 x+1713 x^2
                -2268 x^3+1725 x^4-690 x^5+115 x^6\big) \zeta_3
\Big)
\nn\\
&
-\frac{32 \fxmo \iomx^5}{81} \big(
        48-47 x-186 x^2+250 x^3-114 x^4
        +25 x^5\big) H_{0,1}(x)
-\frac{64}{9} H_{0,0,0,1}(x)
\nn\\
&
-\frac{128}{9} H_{1,0,0,1}(x)
-\frac{128}{45} \iomx (x-9) \zeta_2
-\frac{64}{15} \zeta_2^2
+\frac{128}{9} H_1(x) \zeta_3
\bigg] \,.
%
%
\end{align}

\subsection{High energy expansion of the form factors}
The high energy expansion of the HLFF provides a unique opportunity to study the underlying 
universal structure of QCD scattering amplitudes in terms of various anomalous dimensions and
coefficient functions. 
Such studies were performed for massive form factors (HQFF) in refs.~\cite{Mitov:2006xs,Ahmed:2017gyt,Blumlein:2018tmz}. 
In ref.~\cite{Datta:2023otd}, we presented for the first time the asymptotic behaviour of the HLFF
at three- and four-loop level. 
We showed that in the high energy limit, HLFF exhibit a combined behaviour of the massless
and massive form factors, and hence can be predicted using the ingredients known from these computations.

The predicted results indeed agree with our present explicit computation in the high energy limit,
confirming the correctness of both the present computation and the asymptotic behaviour. 
We note that the predicted results in \cite{Datta:2023otd} do not contain any contributions 
from massive quark loop.
For completeness, we present in the following 
the high energy limit of the present results, i.e. the complete light fermionic 
and double heavy quark loop contributions to $G_1^{(3)}$ and $S^{(3)}$.
The Sudakov logarithm ($L$) in the expressions is defined as followed
\begin{equation}
 L = \ln \left(- \frac{q^2}{m_t^2} \right) \,.
\end{equation}
\begin{align}
 G_{1,\text{asy}}^{(3)} &=
C_F n_l^2 T_F^2 \bigg[
\frac{1}{\ep^4} \bigg\{
-\frac{88}{81}
\bigg\}
+ \frac{1}{\ep^3} \bigg\{
 \frac{32}{27} L
-\frac{160}{243}
\bigg\}
+ \frac{1}{\ep^2} \bigg\{
-\frac{160}{81} L
+\frac{172}{81}
+\frac{8}{9} \zeta_2
\bigg\}
\nn\\&
+ \frac{1}{\ep} \bigg\{
-\frac{32}{81} L
+\frac{5266}{2187}
-\frac{40}{27} \zeta_2
-\frac{16}{81} \zeta_3
\bigg\}
+ \bigg\{
-\frac{8}{27} L^4
+\frac{304}{81} L^3
+L^2 \Big(
        -\frac{1624}{81}
        -\frac{32}{9} \zeta_2
\Big)
\nn\\&
+L \Big(
        \frac{39352}{729}
        +\frac{608}{27} \zeta_2
        +\frac{64}{27} \zeta_3
\Big)
-\frac{322979}{6561}
-\frac{712}{27} \zeta_2
+\frac{1016}{135} \zeta_2^2
+\frac{2624}{243} \zeta_3
\bigg\}
\bigg]
%
\nn\\&
+ C_F^2 n_l T_F \bigg[
\frac{1}{\ep^5} 
+\frac{1}{\ep^4} \bigg\{
-\frac{10}{3} L
+\frac{65}{18}
\bigg\}
+ \frac{1}{\ep^3} \bigg\{
 \frac{11}{3} L^2
-\frac{19}{3} L
+\frac{253}{54}
-\frac{5}{2} \zeta_2 
\bigg\}
\nn\\&
+ \frac{1}{\ep^2} \bigg\{
-\frac{11}{9} L^3
-\frac{37}{18} L^2
+L \Big(
        9
        +\frac{23 \zeta_2}{3}
\Big)
-\frac{746}{81}
-\frac{17}{4} \zeta_2
-9 \zeta_3 
\bigg\}
+ \frac{1}{\ep} \bigg\{
-\frac{25}{36} L^4
+\frac{223}{18} L^3
\nn\\&
-L^2 \Big(
         \frac{2969}{54}
        +\frac{61}{6} \zeta_2
\Big)
+L \Big(
        \frac{1067}{9}
        +\frac{43}{2} \zeta_2
        +\frac{98}{9} \zeta_3
\Big)
-\frac{44411}{486}
-\frac{749}{36} \zeta_2
+\frac{757}{360} \zeta_2^2
+\frac{1241}{54} \zeta_3 
\bigg\}
\nn\\&
+ \bigg\{
 \frac{41}{36} L^5
-\frac{3265}{216} L^4
+L^3 \Big(
        \frac{4987}{54}
        +\frac{149}{18} \zeta_2
\Big)
+L^2 \Big(
        -\frac{53683}{162}
        -\frac{599}{12} \zeta_2
        +29 \zeta_3
\Big)
\nn\\&
+L \Big(
        \frac{32273}{54}
        +\frac{345}{2} \zeta_2
        -\frac{389}{60} \zeta_2^2
        -\frac{2153}{9} \zeta_3
\Big)
-\frac{1132067}{2916}
-\frac{256}{3} a_4
+\Big(
        -\frac{16211}{54}
        +\frac{25}{18} \zeta_3
\nn\\&
        +\frac{224}{3} \log (2)
        -\frac{128}{3} \log ^2(2)
\Big) \zeta_2
+\frac{349843}{2160} \zeta_2^2
+\frac{72127}{162} \zeta_3
+\frac{229}{9} \zeta_5
\bigg\}
\bigg]
%
\nn\\&
+ C_F C_A n_l T_F \bigg[
\frac{1}{\ep^4} \bigg\{
\frac{484}{81} 
\bigg\}
+ \frac{1}{\ep^3} \bigg\{
-\frac{176}{27} L
+\frac{40}{243}
+\frac{20}{27} \zeta_2
\bigg\}
+ \frac{1}{\ep^2} \bigg\{
 L \Big(
        \frac{1336}{81}
        -\frac{16}{9} \zeta_2
\Big)
\nn\\&
-\frac{3856}{243}
-\frac{404}{81} \zeta_2
+\frac{188}{27} \zeta_3
\bigg\}
+ \frac{1}{\ep} \bigg\{
 L \Big(
        -\frac{836}{81}
        +\frac{160}{27} \zeta_2
        -\frac{112}{9} \zeta_3
\Big)
-\frac{3853}{2187}
+\frac{1874}{243} \zeta_2
\nn\\&
+\frac{44}{15} \zeta_2^2
-\frac{220}{81} \zeta_3
\bigg\}
+ \bigg\{
 \frac{44}{27} L^4
+L^3 \Big(
        -\frac{1948}{81}
        +\frac{16}{9} \zeta_2
\Big)
+L^2 \Big(
        \frac{11752}{81}
        +\frac{32}{3} \zeta_2
        -16 \zeta_3
\Big)
\nn\\&
+L \Big(
        -\frac{309838}{729}
        -\frac{11728}{81} \zeta_2
        +\frac{88}{15} \zeta_2^2
        +\frac{1448}{9} \zeta_3
\Big)
+\frac{2866346}{6561}
+\frac{128}{3} a_4
+\Big(
        \frac{129091}{729}
\nn\\&
        +\frac{8}{3} \zeta_3
        -\frac{112}{3} \log (2)
        +\frac{64}{3} \log ^2(2)
\Big) \zeta_2
-\frac{10922}{135} \zeta_2^2
-\frac{28424}{81} \zeta_3
+98 \zeta_5
\bigg\}
\bigg]
%
\nn\\&
+ C_F n_l n_h T_F^2 \bigg[
\frac{1}{\ep^2} \bigg\{
\frac{4}{3} \zeta_2 
\bigg\}
+ \frac{1}{\ep} \bigg\{
-\frac{16}{9} L \zeta_2
+\frac{40}{27} \zeta_2
-\frac{8}{9} \zeta_3
\bigg\}
+ \bigg\{
 \frac{608}{81} L^3
-\frac{16}{27} L^4
-L^2 \Big(
         \frac{3248}{81}
\nn\\&
        +\frac{64}{9} \zeta_2
\Big)
+L \Big(
        \frac{7408}{81}
        +48 \zeta_2
        +\frac{416}{27} \zeta_3
\Big)
-\frac{67636}{729}
-\frac{2602}{81} \zeta_2
-\frac{1121}{45} \zeta_2^2
-\frac{3728}{81} \zeta_3 
\bigg\}
\bigg]
%
\nn\\&
+ C_F n_h^2 T_F^2 \bigg[
 \bigg\{
-\frac{8}{27} L^4
+\frac{304}{81} L^3
+L^2 \Big(
        -\frac{1624}{81}
        -\frac{32}{9} \zeta_2
\Big)
+L \Big(
        \frac{15128}{243}
        +\frac{608}{27} \zeta_2
        -\frac{128}{9} \zeta_3
\Big)
\nn\\&
-\frac{111310}{729}
-\frac{19696}{405} \zeta_2
+\frac{332}{45} \zeta_2^2
+\frac{2528}{27} \zeta_3 
\bigg\}
\bigg] \,.
%
\end{align}
\begin{align}
 S_{\text{asy}}^{(3)} &=
C_F n_l^2 T_F^2 \bigg[
\frac{1}{\ep^4} \bigg\{
-\frac{88}{81}
\bigg\}
+ \frac{1}{\ep^3} \bigg\{
 \frac{32}{27} L
-\frac{160}{243}
\bigg\}
+ \frac{1}{\ep^2} \bigg\{
-\frac{160}{81} L
+\frac{172}{81}
+\frac{8}{9} \zeta_2 
\bigg\}
\nn\\&
+ \frac{1}{\ep} \bigg\{
-\frac{32}{81} L
+\frac{5266}{2187}
-\frac{40}{27} \zeta_2
-\frac{16}{81} \zeta_3
\bigg\}
+ \bigg\{
-\frac{8}{27} L^4
+\frac{160}{81} L^3
+L^2 \Big(
        -\frac{400}{81}
        -\frac{32}{9} \zeta_2
\Big)
\nn\\&
+L \Big(
        \frac{3712}{729}
        +\frac{320}{27} \zeta_2
        +\frac{64}{27} \zeta_3
\Big)
-\frac{59891}{6561}
-\frac{1144}{27} \zeta_2
+\frac{1016}{135} \zeta_2^2
-\frac{2560}{243} \zeta_3
\bigg\}
\bigg]
%
\nn\\&
+ C_F^2 n_l T_F \bigg[
\frac{1}{\ep^5}
+ \frac{1}{\ep^4} \bigg\{
-\frac{10}{3} L
+\frac{65}{18}
\bigg\}
+ \frac{1}{\ep^3} \bigg\{
 \frac{11}{3} L^2
-\frac{10}{3} L
+\frac{145}{54}
-\frac{5}{2} \zeta_2
\bigg\}
+ \frac{1}{\ep^2} \bigg\{
-\frac{11}{9} L^3
\nn\\&
-\frac{50}{9} L^2
+L \Big(
        \frac{29}{3}
        +\frac{23}{3} \zeta_2
\Big)
-\frac{566}{81}
-\frac{21}{4} \zeta_2
-9 \zeta_3
\bigg\}
+ \frac{1}{\ep} \bigg\{
-\frac{25}{36} L^4
+\frac{80}{9} L^3
+L^2 \Big(
        -\frac{971}{54}
\nn\\&
        -\frac{61}{6} \zeta_2
\Big)
+L \Big(
        \frac{529}{18}
        +15 \zeta_2
        +\frac{98}{9} \zeta_3
\Big)
-\frac{16961}{486}
-\frac{1409}{36} \zeta_2
+\frac{757}{360} \zeta_2^2
+\frac{269}{54} \zeta_3 
\bigg\}
+ \bigg\{
 \frac{41}{36} L^5
\nn\\&
-\frac{200}{27} L^4
+L^3 \Big(
        \frac{1213}{54}
        +\frac{149}{18} \zeta_2
\Big)
+L^2 \Big(
        -\frac{19625}{324}
        -\frac{83}{3} \zeta_2
        +29 \zeta_3
\Big)
+L \Big(
        \frac{13309}{108}
        +\frac{809}{6} \zeta_2
\nn\\&
        -\frac{389}{60} \zeta_2^2
        -\frac{1994}{9} \zeta_3
\Big)
-\frac{195257}{2916}
+\frac{256}{3}   a_4
-\Big(
         \frac{343}{27}
        -\frac{25}{18} \zeta_3
        +\frac{1184}{3} \log (2)
        -\frac{128}{3} \log ^2(2)
\Big) \zeta_2
\nn\\&
+\frac{6547 \zeta_2^2}{2160}
+\frac{64171}{162} \zeta_3
+\frac{229}{9} \zeta_5
\bigg\}
\bigg]
%
\nn\\&
+ C_F C_A n_l T_F \bigg[
\frac{1}{\ep^4} \bigg\{
\frac{484}{81}
\bigg\}
+ \frac{1}{\ep^3} \bigg\{
-\frac{176}{27} L
+\frac{40}{243}
+\frac{20}{27} \zeta_2
\bigg\}
+ \frac{1}{\ep^2} \bigg\{
 L \Big(
        \frac{1336}{81}
        -\frac{16}{9} \zeta_2
\Big)
\nn\\&
-\frac{3856}{243}
-\frac{404}{81} \zeta_2
+\frac{188}{27} \zeta_3
\bigg\}
+ \frac{1}{\ep} \bigg\{
 L \Big(
        -\frac{836}{81}
        +\frac{160}{27} \zeta_2
        -\frac{112}{9} \zeta_3
\Big)
-\frac{3853}{2187}
+\frac{1874}{243} \zeta_2
\nn\\&
+\frac{44}{15} \zeta_2^2
-\frac{220}{81} \zeta_3 
\bigg\}
+ \bigg\{
 \frac{44}{27} L^4
+L^3 \Big(
        -\frac{1156}{81}
        +\frac{16}{9} \zeta_2
\Big)
+L^2 \Big(
        \frac{3454}{81}
        +\frac{32}{3} \zeta_2
        -16 \zeta_3
\Big)
\nn\\&
+L \Big(
        -\frac{14998}{729}
        -\frac{6544}{81} \zeta_2
        +\frac{88}{15} \zeta_2^2
        +\frac{1448}{9} \zeta_3
\Big)
+\frac{854630}{6561}
-\frac{128}{3} a_4
+\Big(
        \frac{77899}{729}
        +\frac{8}{3} \zeta_3
\nn\\&
        +\frac{592}{3} \log (2)
        -\frac{64}{3} \log ^2(2)
\Big) \zeta_2
-\frac{9158}{135} \zeta_2^2
-\frac{16652}{81} \zeta_3
+98 \zeta_5 
\bigg\}
\bigg]
%
\nn\\&
+ C_F n_l n_h T_F^2 \bigg[
 \frac{1}{\ep^2} \bigg\{
\frac{4}{3} \zeta_2 
\bigg\}
+ \frac{1}{\ep} \bigg\{
-\frac{16}{9} L \zeta_2
+\frac{40}{27} \zeta_2
-\frac{8}{9} \zeta_3
\bigg\}
+ \bigg\{
-\frac{16}{27} L^4
+\frac{320}{81} L^3
-L^2 \Big(
         \frac{800}{81}
\nn\\&
        +\frac{64}{9} \zeta_2
\Big)
+L \Big(
        -\frac{512}{81}
        +\frac{80}{3} \zeta_2
        +\frac{416}{27} \zeta_3
\Big)
-\frac{51940}{729}
+\frac{6038}{81} \zeta_2
-\frac{1121}{45} \zeta_2^2
-\frac{2000}{81} \zeta_3 
\bigg\}
\bigg]
%
\nn\\&
+ C_F n_h^2 T_F^2 \bigg[
 \bigg\{
-\frac{8}{27} L^4
+\frac{160}{81} L^3
+L^2 \Big(
        -\frac{400}{81}
        -\frac{32}{9} \zeta_2
\Big)
+L \Big(
        \frac{3248}{243}
        +\frac{320}{27} \zeta_2
        -\frac{128}{9} \zeta_3
\Big)
\nn\\&
-\frac{124846}{729}
-\frac{2848}{405} \zeta_2
+\frac{332}{45} \zeta_2^2
+\frac{3680}{27} \zeta_3
\bigg\}
\bigg] \,.
%
\end{align}
\subsection{Checks of the result}
As first check, we have compared numerical evaluation of the analytic expressions of the MIs
with the output of dedicated numerical programs \textsc{FIESTA} and \textsc{AMFlow} for several 
values of $x$, finding excellent agreement.
The fulfillment of the universal behaviour of the IR structure of the renormalized form factors 
acts as another strong check of the computation. 
We have also shown that the Ward identity is satisfied.
Lastly, we find perfect agreement with the results obtained in \cite{Fael:2024vko}
for the relevant color factors.

\section{Conclusions} 
\label{sec:conclusions}
We have presented the analytic results from light-fermionic and double heavy quark 
contributions to the HLFF at three loops in perturbative QCD.
We have followed the traditional methods to compute multiloop scattering amplitudes
by using the method of IBP reduction to reduce the scalar Feynman integrals to MIs
and by using the method of differential equations to solve them.
We renormalize the heavy and light quark fields and the heavy quark mass in the OS scheme,
and the strong coupling constant in $\overline{\text{MS}}$ scheme. 
The UV renormalized form factors satisfy the universal IR structure.
We perform the IR subtraction to obtain the finite remainders. 
The results have been expressed in terms of harmonic polylogarithms 
and generalized harmonic polylogarithms.

\section*{Acknowledgements}
We would like to thank B. Ananthanarayan, J. Bl\"umlein, P. Marquard, V. Ravindran and M. Steinhauser for fruitful discussions.
We thank S. Bera for his support with the series expansion of hypergeometric functions.
We also thank the Centre for High Energy Physics for the help
with computational resources. N.R. extends sincere gratitude to S. Moch for the hospitality at the University of Hamburg, where a part of this work was carried out.
N.R. is partially supported by the SERB-SRG under Grant No. SRG/2023/000591.

\bibliography{main}
\bibliographystyle{JHEP}

\end{document}